\documentclass{optica-article}
\journal{opticajournal} 

\articletype{Research Article}

\begin{document}

\title{Solution for the finite space-bandwidth limitation in digital holography}
\author{Byung Gyu Chae}

\address{Holographic Contents Research Laboratory, Electronics and Telecommunications Research Institute, 218 Gajeong-ro, Yuseong-gu, 
Daejeon 34129, Republic of Korea}

\email{bgchae@etri.re.kr}

\begin{abstract}

A lensless digital holography enables wide-field microscopic imaging without the limitations imposed by optical lens performance.
However, conventional holographic imaging often relies on magnifying optical systems to compensate for the low resolution of holograms captured by image sensors.
The spatial resolution of the reconstructed image is fundamentally constrained by the space-bandwidth of the hologram due to aliasing errors at insufficient sampling rates.
This study analyzes the spatial distribution of the angular spectrum in undersampled holograms using angle modulation techniques.
Aliased replica functions are identified as phase-modulated functions by multiples of the sampling frequency,
with the spatial frequency components continuously extending into the replica regions.
Optical imaging simulations demonstrate that image reconstruction beyond the space-bandwidth limitation of digital holograms is feasible.
In particular, high-order diffraction fields, characterized by orthogonality, can be effectively eliminated through an upsampling process.
By sequentially removing high-order terms and applying a learning-based denoising algorithm, wide-field high-resolution optical imaging is achieved.
This approach demonstrates that only a captured low-resolution hologram can reconstruct a high-resolution image,
thereby overcoming the limitations imposed by the finite space-bandwidth of digital holography.
\end{abstract}


\section{Introduction}

Digital holography is a technique that carries out the imaging using the optically captured diffractive wave \cite{1,2}.
This method makes it possible to acquire the optical image without the help of magnifying optical lens system \cite{3,4,5,6}.
In conventional microscopy, the optical lens limits the range of the angular spectrum of the diffracted wave,
requiring lenses with higher numerical apertures (NA) to achieve high-resolution imaging.
However, these high-NA lens systems typically capture high-resolution images only within a submillimeter field of view to mitigate optical aberrations,
which is an inherent bottleneck in realizing wide-field imaging systems \cite{7,8,9,10,11}.
In practice, digital holography still often relies on magnifying optics to visualize subcellular structures in biomedical imaging or fine material details, 
due to the microscale pixel size of image sensors.
While high-resolution holograms can be obtained using nanoscale sensor arrays or pixel super-resolution techniques,
such technologies are currently limited in availability or involve complex implementations.

Previous studies have shown that high-resolution images can be reconstructed from low-resolution holograms captured by lensless digital holography \cite{12}.
The imaging performance depends on how much of the higher-frequency components of the angular spectrum of the diffracted waves is collected.
A digital hologram records the diffracted waves in a finite digitized area, and its space-bandwidth determines the spatial resolution of the reconstructed image.
The Fresnel hologram sampled at $\Delta$ pixel interval with $N$$\times$$N$ pixels has the space-bandwidth $B_w$ \cite{13}:
\begin{equation}
B_w=\frac{N\Delta}{\lambda z}. 
\end{equation}
During the imaging process, the angular spectrum of the digital hologram evolves with propagation distance in free space,
changing only in the relative phases of its spectral components \cite{1,14}.
Thus, the resolution limit $R_{\rm{lim}}$ of the reconstructed image is directly given as the reciprocal of bandwidth \cite{15,16,17,18},
$R_{\rm{lim}}= \frac{\lambda z}{N\Delta}=\frac{\lambda}{2 \rm{NA}}$.
The NA of hologram is $\sin(\it{\Omega}/\rm{2})$, according to the Abbe sine condition.

Meanwhile, aliased replica patterns are formed when the required bandwidth exceeds the bandwidth supported by the hologram's pixel \cite{19,20,21,22,23}.
Spatial frequency components higher than the Nyquist frequency are aliased, making them apper as lower frequency components of the original function.
The spatial frequency $f$ increases linearly along the spatial direction, and the aliased frequency $f_a$ is determined by the relationship \cite{24}:
\begin{equation}
f_a = |f-nf_s|,
\end{equation}
where $f_s$ is the sampling frequency and $n$ is an integer. 
This property results in the formation of replica functions that are spatially distributed at regular periodic intervals.

\begin{figure}[ht!]
\includegraphics[scale=0.8, trim= 1.3cm 13.7cm 0cm 0.5cm]{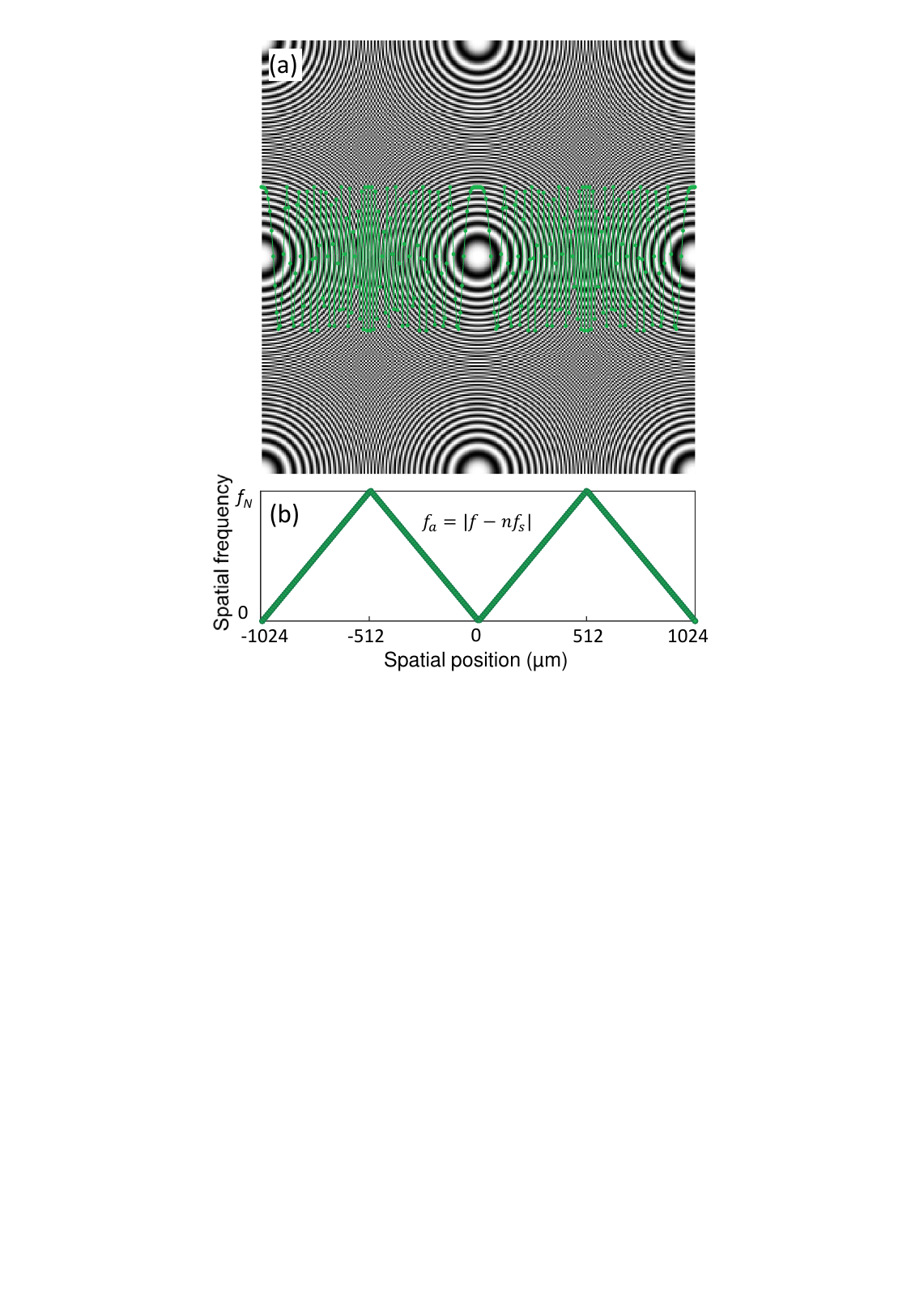}
\caption{Aliasing phenomenon for a point-source hologram sampled at a lower sampling rate.
Digital hologram consisting of 256$\times$256 pixels with a 8-$\mu$m pixel pitch was synthesized at a distance of a half of $z_c$.
(a) The real-valued hologram exhibits four replica Fresnel zones.
The 1D quadratic phase function is overlapped along the lateral distance at the center.
(b) Aliased frequencies for the quadratic phase function.}
\end{figure}

Figure 1 illustrates the aliasing phenomenon for a point-source hologram sampled at a lower sampling rate.
The complex-valued hologram $g(x,y)$ for a point object is expressed as the impulse response function,
$ g(x,y) = \frac{e^{i kz}}{i\lambda z} \exp \left[\frac{i\pi}{\lambda z} (x^2 + y^2) \right] $.
Using the shifted form of the Fourier-transformed function, the sampled hologram field at an interval $\Delta x$ along the $x$-axis is written as follows \cite{25,26,27}:
\begin{equation}
\sum_{n} g(n \Delta x) \delta(x-n \Delta x) = \frac{1}{\Delta x} \sum_{n} c_n g \left( x + \frac{\lambda zn}{\Delta x} \right).
\end{equation}
For simplicity, one-dimensional description is used hereafter.
The angular spectra, within the range of the original function, are folded at intervals determined by the folding frequency.
Continous response functions are obtained through the inverse Fourier transform of shifted angular spectra.
The replication patterns are formed at a reduced period of $\frac{\lambda z}{s \Delta x}$ when undersampled by a factor $s$ of $\Delta x$.

The explanation of aliasing in digital holography leaves it unclear whether a holographic image can be reconstructed with a space-bandwidth that fully spans the aperture of the digital hologram,
based on Eq. (1).
However, both numerical and experimental studies suggest that the replica functions in undersampled holograms correspond to high spatial-frequency components of the hologram field
\cite{12,26,27}.
Despite this, the space-bandwidth seems to be confined to that of the original function due to replication, meaning the diffraction performance is limited to its initial capacity.
Consequently, establishing a unified understanding of the role and utility of these replica functions has been challenging.
If it can be clearly demonstrated that aliased fringes contain true high-frequency information, then wide-field high-resolution imaging could become feasible without the need for optical magnification.

In this study,
an angular spectrum distribution in the undersampled hologram with aliased replica fringes is first investigated by applying angle modulation in the complex plane.
The replica functions are found to display phase modulation patterns formed by carrier waves corresponding to integer multiples of the sampling frequency.
The spatial frequency increases linearly across the replica regions, 
indicating that these replicas encode higher frequency components of the original signal, rather than serving merely as duplicated copies.
Second, numerical simulations of optical imaging validate this interpretation and further demonstrate that
high-order diffraction components can be selectively suppressed by exploiting their orthogonal properties.
Based on these findings, a method for wide-field high-resolution optical imaging that does not rely on optical magnification is proposed.

\section{Angular spectrum distribution in the undersampled hologram}
\subsection{Spatial frequency in complex plane}

The spatial frequency in the Fresnel hologram is represented as $\frac{x}{\lambda z}$, which depends linearly on the lateral spatial distance \cite{1}.
The distribution of spatial frequency in the undersampled hologram of a point object $\delta (x',y')$ is investigated, particularly in the complex plane.
The Fresnel hologram is a linear combination of unit hologram of a point object.

The complex-valued hologram $g(x,y)$, consisting of 256$\times$256 pixels with a 8-$\mu$m pixel pitch, was synthesized using the Fresnel diffraction formula with a single Fourier transform.
A unit-amplitude plane wave with a wavelength $\lambda$ of 532 nm was used.
The point-source hologram placed at a distance of a half of $z_c$ results in the formation of four replica Fresnel zones, as shown in Fig 1(a).
Considering the replication interval $\frac{\lambda z}{\Delta x}$ equal to the field extent $N \Delta x$ in Eq. (3),
the critical distance $z_c$ is defined as the minimum distance at which hologram functions are free from replication effects:
\begin{equation}
z_c = \frac{N {\Delta x}^2}{\lambda}.
\end{equation}
No aliased fringes appear for distances above $z_c$.
For this configuration, $z_c$ is calculated to be 30.8 mm. 

\begin{figure}[ht!]
\includegraphics[scale=0.8, trim= 1.6cm 19.3cm 0cm 0cm]{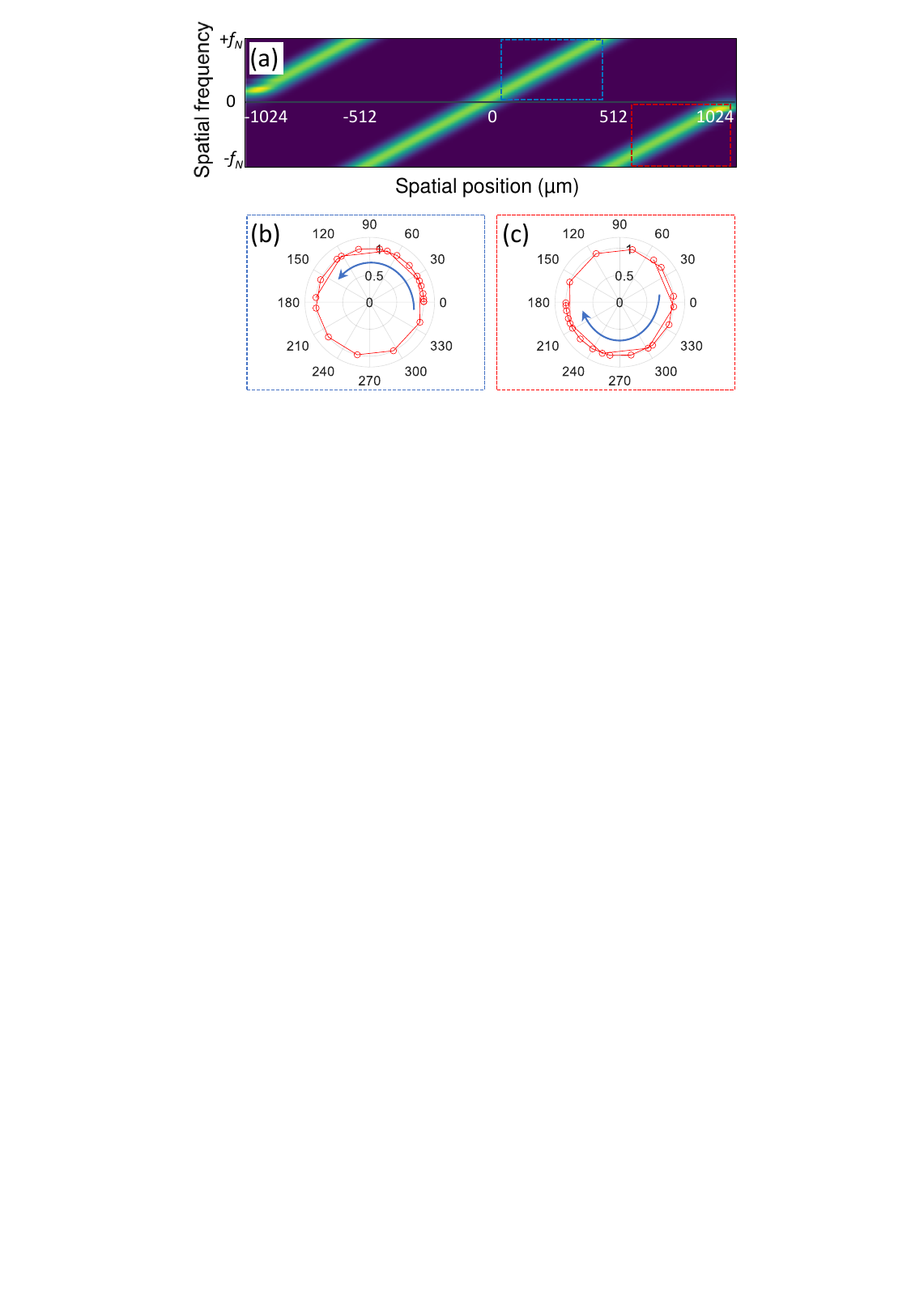}
\caption{Spatial frequency distribution of an undersampled point-source hologram in the complex domain.
(a) Spatial frequency variation along the lateral spatial position.
The trajectories of the position vectors (b) in the blue box and (c) in the red box, corresponding to those in Fig. 2(a), are plotted in polar coordinates, respectively.}
\end{figure}

\begin{figure}[ht!]
\includegraphics[scale=0.8, trim= 1.1cm 15.5cm 0cm 0.2cm]{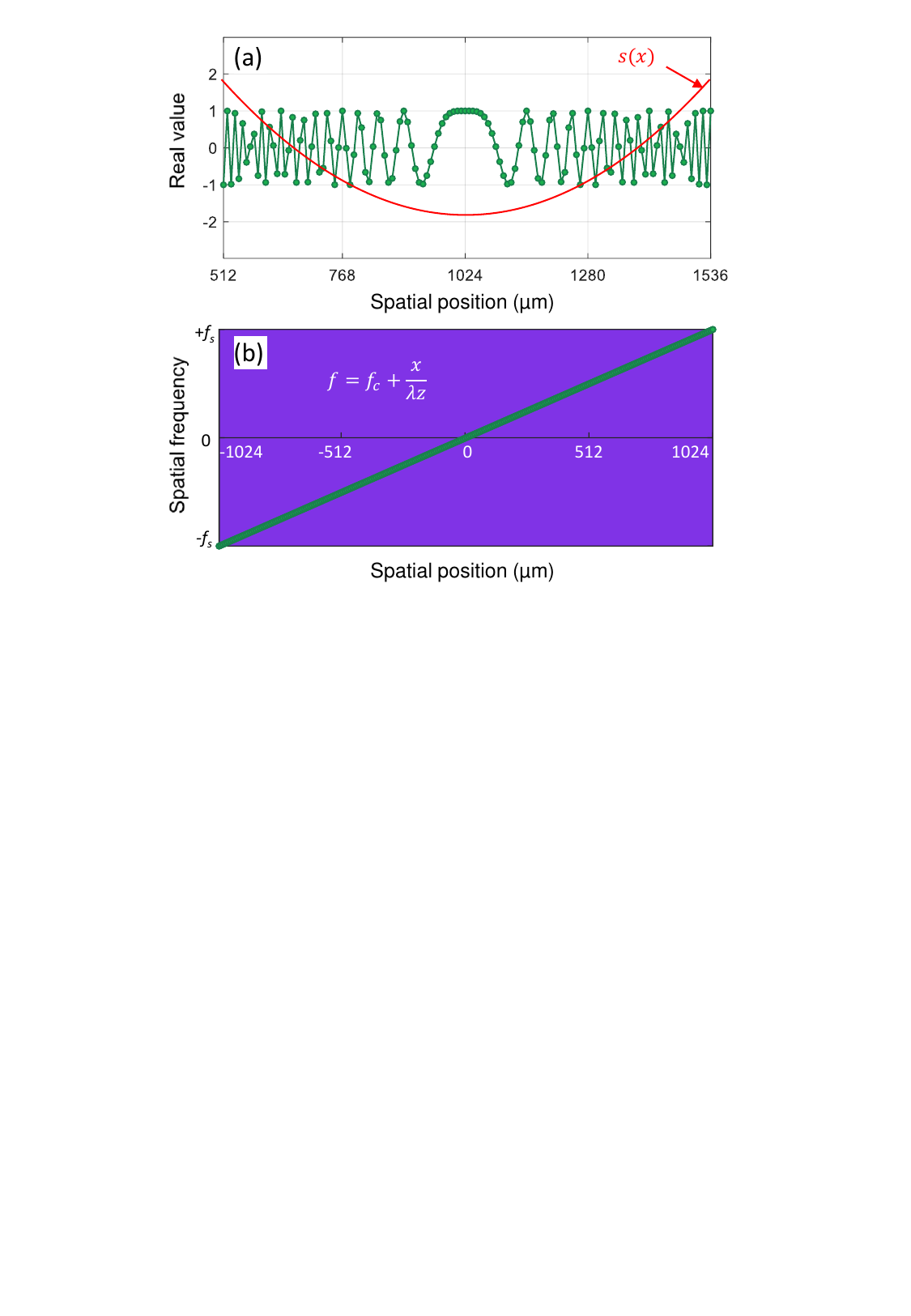}
\caption{ Spatial frequency distribution after considering modulation form by a carrier wave.
(a) The phase profile of the carrier wave modulated by the modulating signal at the +1$\textsuperscript{st}$-order shifted position.
The modulating signal is inserted at an arbitrary position.
(b) The spatial frequency as a function of spatial position.}
\end{figure}

Figure 2 illustrates the change in spatial frequency along the lateral distance.
The spatial frequency along the $x$-axis at the center of the hologram is obtained using a commercial spectrum calculation program, in Fig. 2(a).
Unlike the aliased frequencies in the real domain, both negative and positive spatial frequency components appear in the complex plane.
In the complex-valued function, the position vector with negative frequencies rotates clockwise, while the position vector with positive frequencies rotates counterclockwise.

The spatial frequency exhibits a linear dependence on the lateral spatial distance.
On the positive-axis side, the spatial frequency increases continuously up to the Nyquist frequency, i.e., $f_N = \frac{1}{2} \left( \frac{1}{\Delta x} \right) = 6.25\times10^4 \rm{m}^{-1}$.
The position of the Nyquist frequency is 512 $\mu$m.
At this point, the spatial frequency changes discontinuously to the negative Nyquist value, then continues to increse toward zero.
The trajectories of the position vectors in these two distinct regions are plotted in polar coordinate, as depicted in Figs. 2(b) and 2(c).
The position vector rotates from a counterclockwise direction to a clockwise direction at the Nyquist frequency. 
Similarily, the negative side shows symmetrical behavior.

\subsection{Angle modulation in the undersampled hologram}

The spatial frequency distribution in an undersampled hologram can be interpreted using angle modulation in the complex plane.
The shifted replica function in Eq. (3) is described as a phase-modulated function by a carrier wave: 
\begin{equation}
c_n g \left( x + \frac{\lambda zn}{\Delta x} \right) = \exp \left[i2\pi \left(\frac{nx}{\Delta x} + \frac{x^2}{2\lambda z} \right) \right].
\end{equation}
In the phase modulation \cite{24}, the instantaneous phase $\phi(x)$ is given by
\begin{equation}
\phi(x) = 2\pi f_c x + D_o s(x),
\end{equation}
where the carrier wave is a sinusoidal wave with a carrier frequency, $f_c = n\left(\frac{1}{\Delta x}\right)$, expressed as $e^{\frac{-i2\pi n x} {\Delta x}}$,
and the modulation index $D_o$ is equal to one.

The modulating signal $s(x)$ in the above equation is a simple quadratic function:
\begin{equation}
s(x) = \frac{\pi x^2}{\lambda z}.
\end{equation}
The phase of the carrier wave is modulated by the modulating signal,
leading to the formation of a replica zone, which appears as moire fringes shifted to $\frac{\lambda zn}{\Delta x}$.
Figure 3(a) shows the +1$\textsuperscript{st}$-order replica function shifted to a position of 1024 $\mu$m.
Since the amplitude of the phase-modulated wave remains constant, the average power of the spatially distributed wave is one half.

The instantaneous frequency in phase modulation is calculated from the derivative of phase value:
\begin{equation}
f = \frac{1}{2\pi} \frac{d \phi(x)}{dx} = f_c + D_o \frac{ds(x)}{dx} = f_c + \frac{x}{\lambda z}.
\end{equation}
The spatial frequency $f$ is the sum of the original frequency $\frac{x}{\lambda z}$ and the carrier frequency.
Here, the carrier frequency is a multiple of the sampling frequency.
Figure 3(b) draws the angular spectrum distribution considering the modulation behaviour.
The spatial frequency on both sides of the $x$-axis increases from zero to the sampling frequency of $f_s = 1.25\times10^5 \rm{m}^{-1}$ without any discontinuity.

Likewise, all shifted replica functions can be expressed as higher spatial-frequency components corresponding to those of the original function.
This analysis in the complex plane demonstrates that the shifted replica function represents not only aliased fringes but also higher spectral components.
Although the description focuses on the central function, it applies equally to the adjacent functions.
It is interesting because the spatial frequency is relative depending on the criteria.
Consequently, the spatial-frequency components of each function should be still the high-frequency components of adjacent functions \cite{26}.

\begin{figure}[ht!]
\includegraphics[scale=0.75, trim= 0.5cm 19cm 0.0cm 0.0cm]{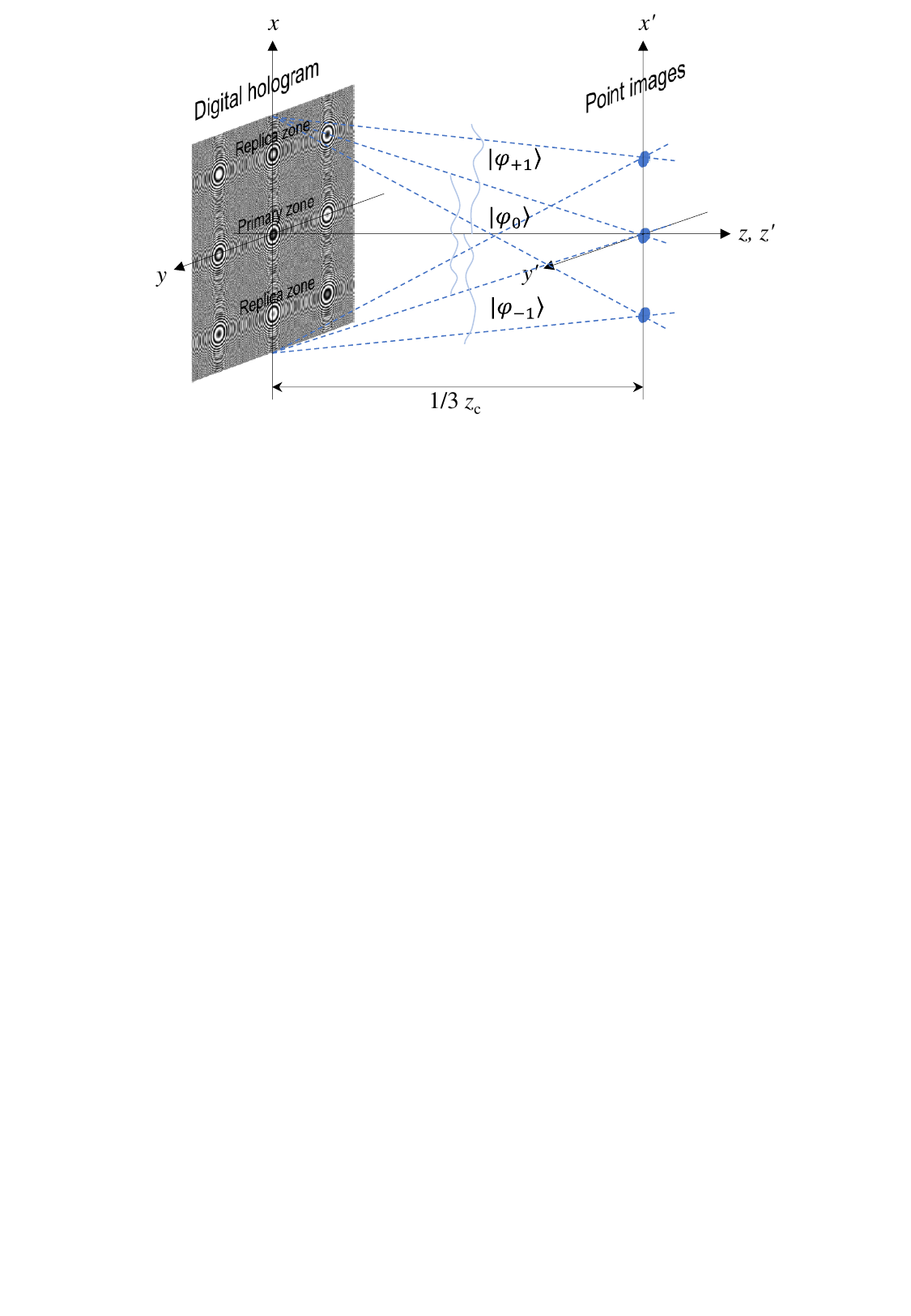}
\caption{Schematic diagram illustrating high-order diffractions from an undersampled hologram.
Point-image reconstruction is described using a hologram made at one-third of $z_c$.
For clarity, high-order diffractions for three Fresnel zones in the central region are drawn.}
\end{figure}

\section{High-order diffraction fields in undersampled holograms}
\subsection{Orthogonality of high-order diffraction fields}

The plane wave incident on an undersampled hologram generates high-order images in the image plane, as illustrated in Fig. 4. 
For clarity, the high-order diffractions for three Fresnel zones, created at one-third of $z_c$, are depicted.
Evidently, each image is reconstructed using the entire aperture of the digital hologram.

The diffraction function $| \psi \rangle$ for the undersampled hologram consists of spatial image states $| \varphi \rangle$ corresponding to individual Fresnelets: 
\begin{equation}
\psi = \sum_{q} \alpha_q | \varphi_q \rangle,
\end{equation}
where $\alpha_q$ represents the probability coefficient.
These coefficients are equal from the fact that point images exhibit uniformly distributed intensity.
It is important that the spatial image states are eigenfunctions, as any function can be expressed as a complete set formed by their linear combination.
The eigenfunctions are orthogonal, verified by the inner product of spatial image states:
\begin{equation}
\langle \varphi_m | \varphi^*_n \rangle = \iint_{-\infty}^{\infty} \varphi_m (x',y') \varphi^*_n (x',y') dx'dy'.
\end{equation}

The spatial image state can be calculated using the Fresnel diffraction formula.
In a one-dimensional description, $\varphi_m (x')$ propagating to an arbitrary distance $z_1$ is given as
\begin{equation}
\varphi_n(x') = \frac{e^{-i kz_1}}{i\lambda z_1} \exp \left(-i\pi\frac{x'^2}{\lambda z_1} \right) \iint g_n(x) \exp \left(-i\pi\frac{x^2}{\lambda z_1} \right) \exp \left(2i\pi\frac{x'x}{\lambda z_1} \right) dx.
\end{equation}
Substituting $g_n(x)=\exp \left[i\pi \left(\frac{x^2}{\lambda z} + \frac{nx}{\Delta x} \right) \right]$ from Eq. (5), above equation becomes
\begin{equation}
\varphi_n(x') = \frac{e^{-i kz_1}}{i\lambda z_1} \exp \left(-i\pi\frac{x'^2}{\lambda z_1} \right) \exp \left(-i\pi\frac{b^2}{a \lambda^2 z_1^2} \right) \iint \exp \left[i\pi a\left(x + \frac{b}{a\lambda z_1} \right)^2 \right] dx,
\end{equation}
where $a = \frac{1}{\lambda z} - \frac{1}{\lambda z_1}$ and $b = x' + \frac{\lambda z_1 n}{2\Delta x}$.
The integral evaluates to $\sqrt{\frac{i}{a}}$ using contour integration:
\begin{equation}
\varphi_n(x') = \frac{e^{-i kz_1}}{i\lambda z_1} \sqrt{\frac{i}{a}} \exp \left(-i\pi\frac{x'^2}{\lambda z_1} \right) \exp \left[-i\pi \frac{1}{a \lambda^2 z_1^2}\left(x' + \frac{\lambda z_1 n}{2\Delta x} \right)^2 \right].
\end{equation}

The inner product of two state functions, $\int_{-\infty}^{\infty} \varphi_m (x') \varphi^*_n (x') dx'$, simplifies to a delta function:
\begin{equation}
\frac{1}{\lambda^2 z_1^2 a} \exp \left[-i\pi \frac{m^2-n^2}{4a\Delta x^2} \right] \int_{-\infty}^{\infty} \exp \left[2i\pi \frac{x'(m-n)}{a \lambda z_1 2\Delta x} \right] dx' 
= \frac{2\Delta x}{\lambda z_1} \delta(m-n).
\end{equation}
Extending this to two-dimensional space yields
\begin{equation}
\iint_{-\infty}^{\infty} \varphi_m (x',y') \varphi^*_n (x',y') dx'dy' = c_o^2 \delta_{mn},
\end{equation}
where $c_o = \frac{2\Delta x}{\lambda z_1}$.
This result confirms that spatial image modes are orthogonal and serve as eigenstates of the diffraction function. 
Additionally, the image intensity is inversely proportional to the square of the propagation distance.

The diffraction fields from individual Fresnelets do not interfere with each other during wave propagation. 
Each diffraction beam independently forms the corresponding image without disturbance from other beams.

This interpretation applies to high-order diffractions generated from the properly sampled hologram. 
This orthogonal property enables the effective removal of high-order terms, commonly treated as noise, using appropriate external operations \cite{27}.

\begin{figure}[ht!]
\includegraphics[scale=0.8, trim= 1.6cm 15.3cm 0cm 0cm]{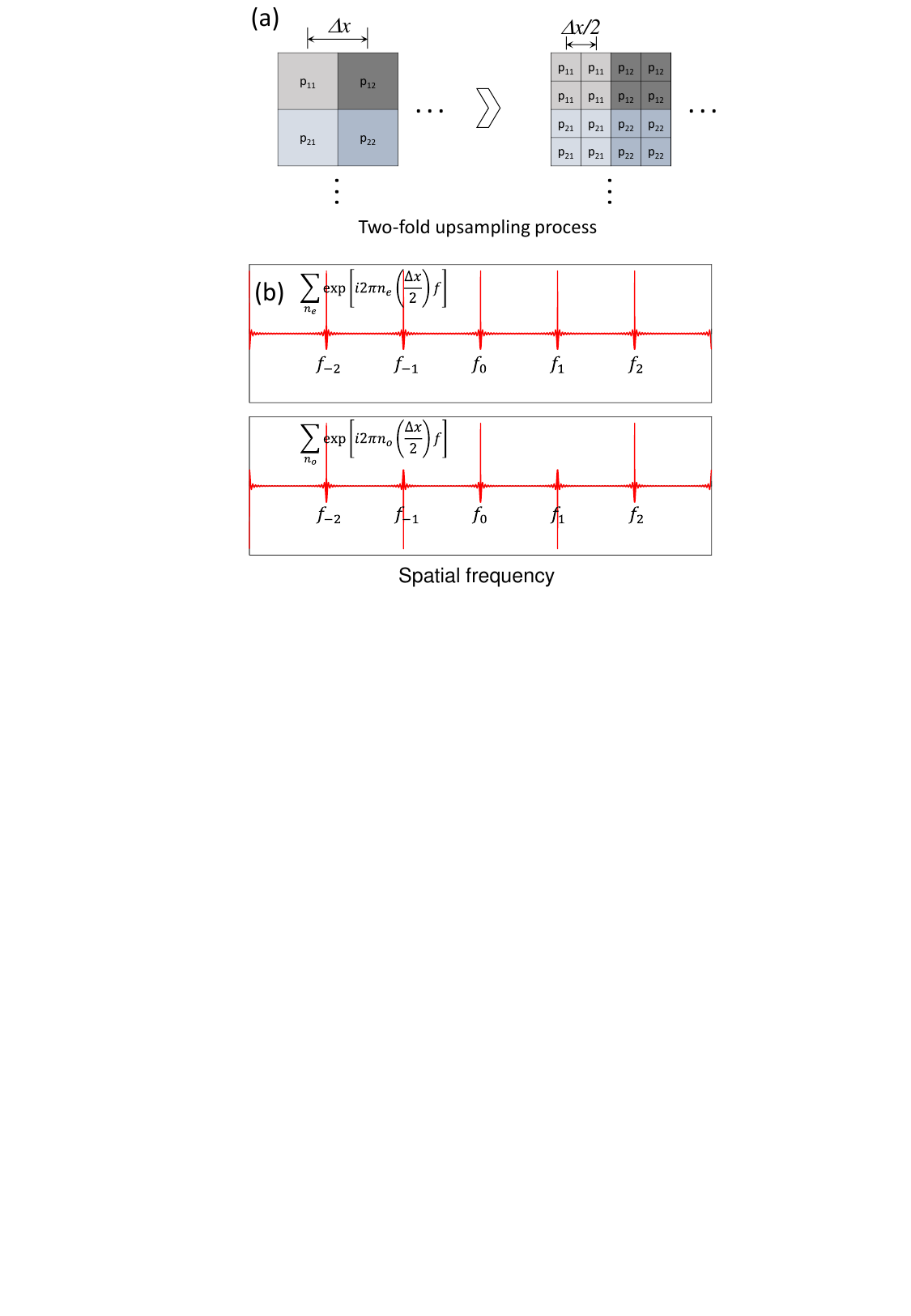}
\caption{Diffraction properties of an upsampled hologram. 
(a) The upsampling process of the sampled hologram.
The two-fold upsampling is achieved by duplicating the pixel values of the original sampled hologram, where each pixel is subdivided into smaller sub-pixels, $p_{mn}$.
(b) The Fourier spectra of Dirac comb functions corresponding to the sampling by even numbers $n_e$ and odd numbers $n_o$ are numerically calculated.
The period of spatial frequencies is a multiple of $\frac{1}{\Delta x}$.}
\end{figure}

\subsection{Suppression of high-order diffraction fields}

The sampled hologram generates high-order diffractions due to its pixelated structure.
The high-order terms in the Fresnel diffraction regime are characterized in terms of a periodic Fourier spectrum $G$ \cite{17}:
\begin{equation}
\textbf{\textit{FT}} \left[ \sum_{n} g(x) \delta(x-n\Delta x) \right] = G(f) \otimes \textbf{\textit{FT}} \left[ \sum_{n} \delta(x-n\Delta x) \right] 
= \frac{1}{\Delta x} \sum_{q} G \left( f - \frac{q}{\Delta x} \right),
\end{equation}
where $\otimes$ represents the convolution sign.
The periodic Fourier spectrum is expressed using the Dirac comb distribution.

Figure 5 illustrates the diffraction properties of sampled hologram using a two-fold upsampling process.
The upsampling process is carried out by duplicating the pixel values of the original hologram, 
where each pixel in the original is subdivided into four identical pixels in the upsampled version.
There are no any interpolation techniques in this upsampling process.

In this circumstance,
the two-fold upsampled hologram can be classified into two components: one sampled by even numbers $n_e$ and the other by odd numbers $n_o$, i.e.,
$\sum_{n_e} \delta \left(x-\frac{n_e \Delta x}{2} \right) + \sum_{n_o} \delta \left(x-\frac{n_o \Delta x}{2} \right)$.
The corresponding pixels in both terms have the same value, but the odd term pixels are shifted by $\frac{\Delta x}{2}$.

Applying this description to the Fourier transform of the Dirac comb distribution in Eq. (16), the function is written by
\begin{equation}
\textbf{\textit{FT}} \left[ \sum_{n} \delta \left(x-\frac{n \Delta x}{2} \right) \right]  = \sum_{n_e} \exp \left[ i2\pi n_e \left(\frac{\Delta x}{2} \right) f \right]
+ \sum_{n_o} \exp \left[ i2\pi n_o \left(\frac{\Delta x}{2} \right) f \right],
\end{equation}
where a Dirac comb in frequency domain is expressed as a Fourier series.
As shown in Fig. 5(b), Dirac delta functions in both terms have an opposite value at odd multiples of $\frac{1}{\Delta x}$,
which is due to the shifted phase factor of $e^{2\pi i \left(\frac{\Delta x}{2} \right) f}$.
Thus, both Fourier transform terms combine into a single function, $\sum_{q} \delta \left( f - \frac{2q}{\Delta x} \right)$,
resulting in the removal of the $\pm$1$\textsuperscript{st}$-order peaks.
The periodic Fourier spectrum of sampled hologram is then represented as
\begin{equation}
\textbf{\textit{FT}} \left[ ... \right] = \frac{2}{\Delta x} \sum_{q} G \left( f - \frac{2q}{\Delta x} \right).
\end{equation}
Only the high-order spectra are generated at intervals of multiples of $\frac{2}{\Delta x}$.
The Fourier space is expanded into two times.
This type of upsampling process effectively suppresses corresponding high-order diffractive waves \cite{17,26,27}.

\begin{figure}[ht!]
\includegraphics[scale=0.75, trim= 0.9cm 17cm 0cm 0cm]{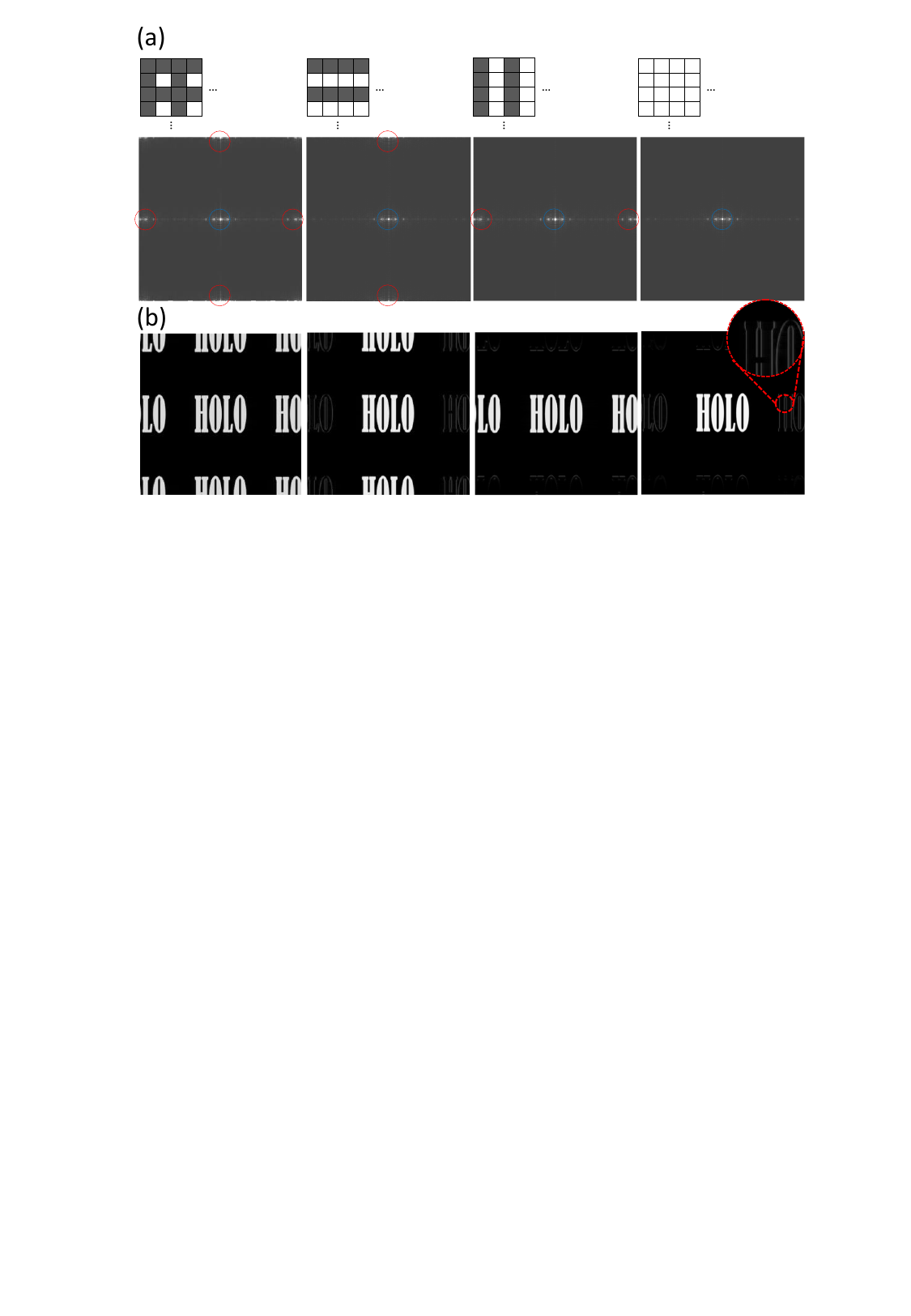}
\caption{Numerical simulation for the suppression of high-order images.
(a) Sub-pixel structure of the 2-fold upsampled hologram and its corresponding Fourier-transformed data.
(b) Reconstructed images from the 2-fold upsampled holograms.
The magnified image reveals remnants of high-frequency components from the high-order images.}
\end{figure}

Figure 6 presents a numerical simulation demonstrating the suppression of high-order diffraction images.
The simulation was performed using the Riemann integral of the diffraction formula, with parameters matching those in Subsection 2.1, except that the image space was expanded twofold.
A digital hologram of the word “HOLO” was synthesized at half the critical distance $z_c$.
In Figure 6(a), the $\pm$1$\textsuperscript{st}$-order peaks in the Fourier-transformed hologram data are eliminated by combining both the even and odd pixel components along each axis.
This upsampling process effectively suppresses the retrieval of the $\pm$1$\textsuperscript{st}$-order diffraction images.
However, as shown in the magnified region of Figure 6(b), residual high-frequency components from the higher-order terms remain visible.
This indicates that the upsampling technique, while effective in suppressing the main high-order replicas, 
still has limitations in completely eliminating all high-frequency artifacts.

\begin{figure}[ht!]
\includegraphics[scale=0.8, trim= 1.7cm 15.7cm 0cm 0cm]{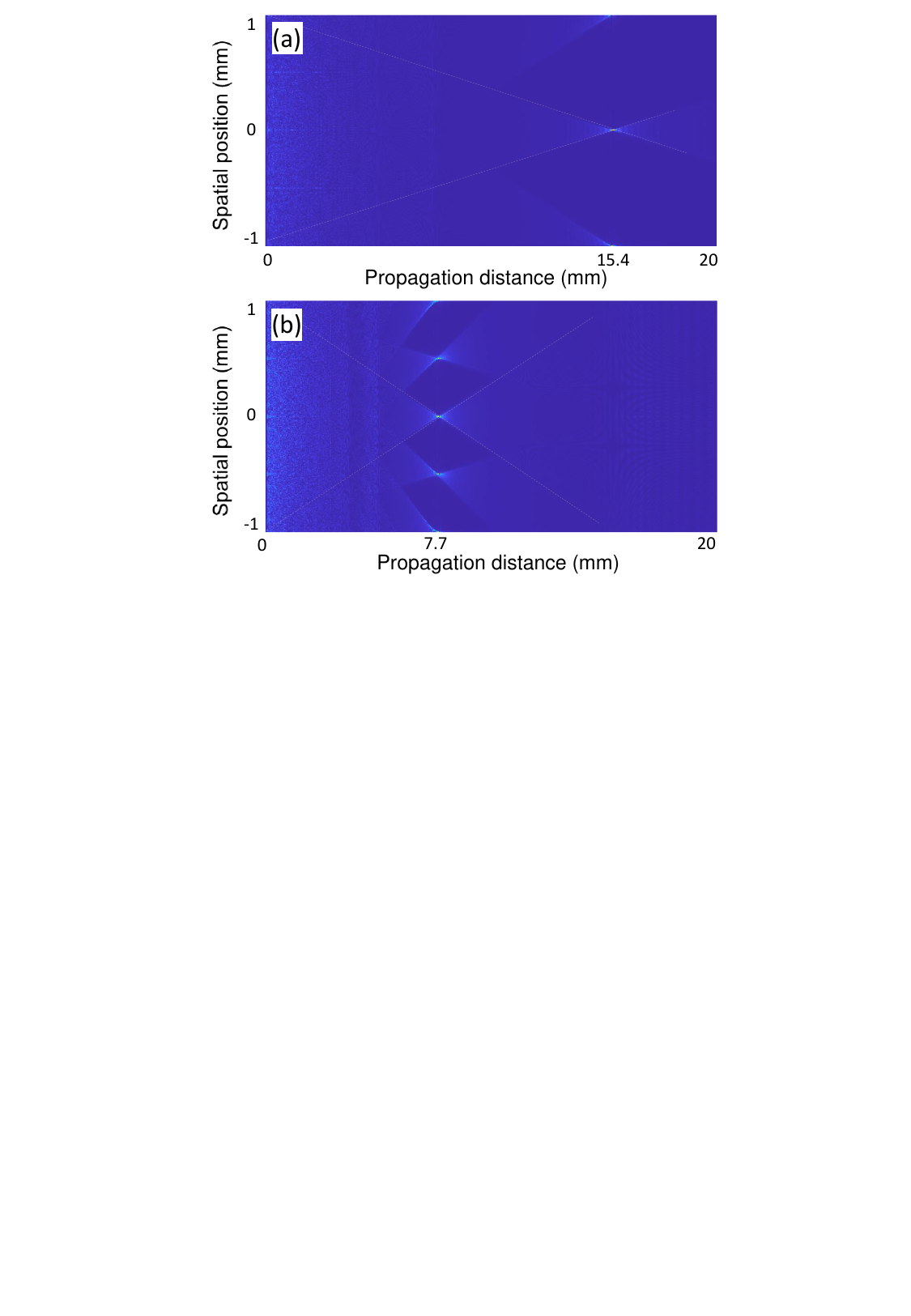}
\caption{Imaging behavior of an undersampled point-source hologram simulated using the Rayleigh-Sommerfeld diffraction formula.
(a) Propagating waves from the hologram prepared using a point object placed at half-$z_c$.
(b) Propagating waves from the hologram prepared using a point object placed at quarter-$z_c$.
The intensity of the diffractive waves along the $x$-axis is displayed.
The aspect ratio of the image is conveniently adjusted.
Inset lines are included to clearly indicate the diffraction angle, $2\theta$.}
\end{figure}

\section{Imaging behaviour of undersampled hologram}

Figure 7 illustrates the imaging behavior for the undersampled hologram.
The diffractive wave $\psi(x',y')$, propagating from the hologram function $g(x,y)$, was numerically calculated by using the Rayleigh-Sommerfeld diffraction formula \cite{1}:
\begin{equation}
\psi(x',y') = \frac{i}{\lambda} \iint E_R(x,y) g(x,y) \frac{\exp (-ikr)}{r} dxdy,
\end{equation}
where an incident plane wave with a unit amplitude, $E_R(x,y)$ is assumed, and $k$ is the wavenumber of $2\pi/\lambda$.
The intensity of the diffractive waves along the lateral line at the center is displayed as a function of the propagating distance, $r=\sqrt{z^2 + (x'-x)^2 + (y'-y)^2}$.
The well-behaved waves propagate at all propagation distances without disturbance.
Individual diffraction waves cannot be distinguished in the region close to the hologram.
However, as previously stated,
high-order diffractions do not interfere with each other because of their orthogonality,
which results in the formation of the point images in the Fresnel diffraction region.
 
Figure 7(a) shows propagating waves from the point-source hologram synthesized at a half of $z_c$.
The proper sampling without overlapping of  the $\pm1 \textsuperscript{st}$-order diffractions appears in the region from 10.3 mm to 30.8 mm,
where the $S$ ratio ranges from 0.67 to 2 (see Section 1 of \textcolor{blue}{Supplement 1}).
These values well match with those in the numerical simulation.
The point image is formed at a half of $z_c$, i.e., 15.4 mm, and then spreads out as it propagates further.
The image resolution of 4 $\mu$m is twice as high as the hologram pixel size,
and similarly, the spreading angle is double that of the diffraction angle of hologram pixel.
The spreading angle, i.e., the viewing angle $\it{\Omega}$ is estimated to be approximately 7.6$^\circ$,
whereas the diffraction angle $\theta$ for a hologram pixel size of 8 $\mu$m is 3.8$^\circ$.

The hologram made at a quarter of critical distance reconstructs the point image with a spatial resolution of 2 $\mu$m,
where the viewing angle increases to 15.2$^\circ$, in Fig. 7(b).
The proper sampling area ranges from 6.2 mm to 10.3 mm, which agrees with the calculated quantities.

Both the image resolution and spreading angle increase as the synthesis distance of the digital hologram decreases.
It appears that the entire aperture of the digital hologram contributes to image formation, regardless of aliasing.

\begin{figure}[ht!]
\includegraphics[scale=0.8, trim= 0.7cm 15.5cm 0cm 0cm]{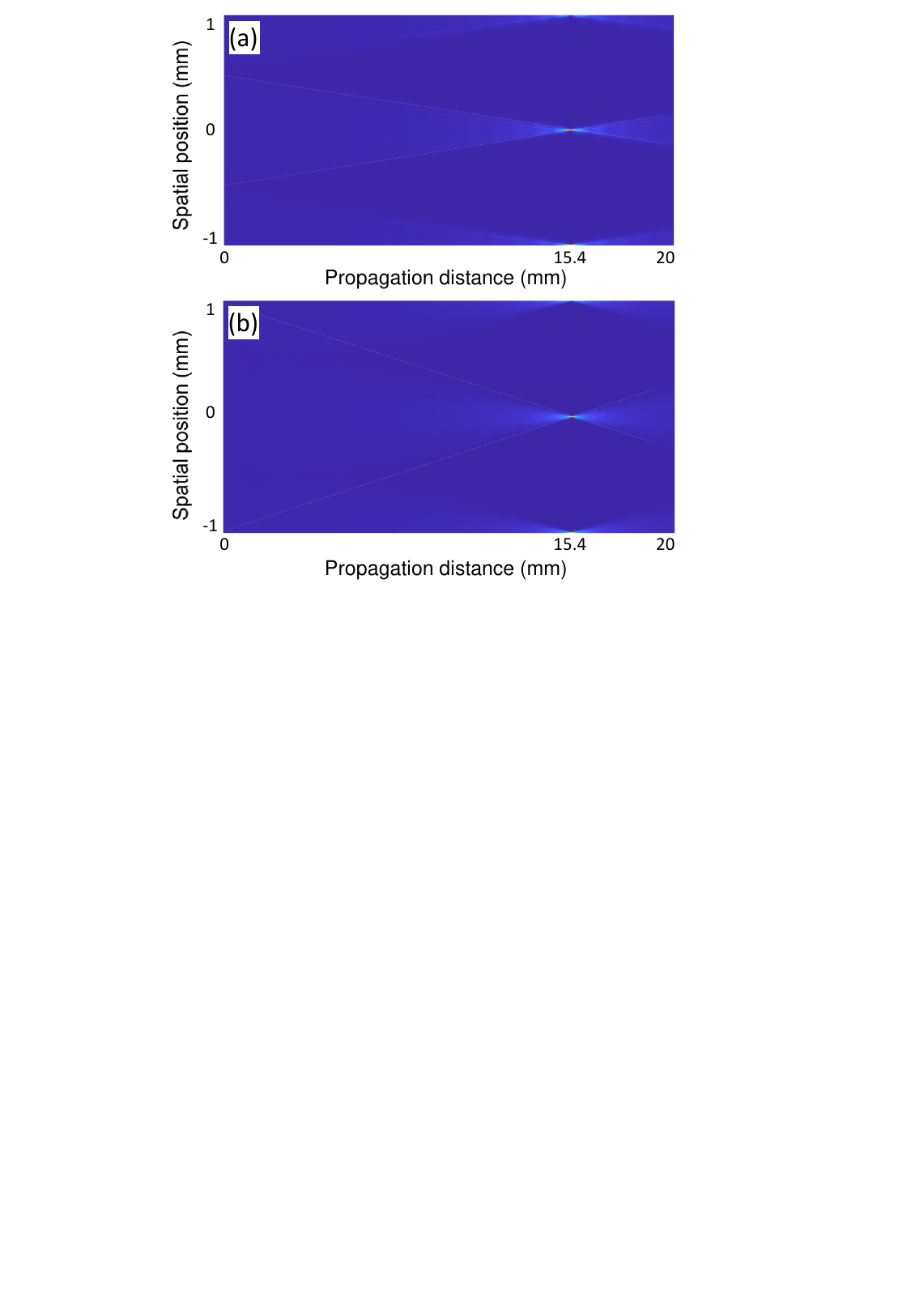}
\caption{Simulated propagating waves using the angular spectrum method.
(a) High-frequency components of diffractions are eliminated and the point image is formed from the primary zone.
(b) The diffraction behavior of two-fold upsampled hologram is illustrated.}
\end{figure}

Meanwhile,
the angular spectrum method allows the image space to be confined within the diffraction zone of a hologram pixel,
effectively eliminating overlapping of high-order diffractions during numerical calculations.
The algorithm involves a double Fourier transform \cite{1,2}: 
\begin{equation}
\psi(x',y') = \textbf{\textit{IFT}} \left( \textbf{\textit{FT}} \left[ E_{R}(x,y) g(x,y) \right]  \exp \left[ i\pi \lambda z (f_x^2 + f_y^2)  \right] \right),
\end{equation}
where $\textbf{\textit{FT}}$ and $\textbf{\textit{IFT}}$ indicate the Fourier transform and its inverse, respectively.
The pixel resolutions in each space are governed by the sampling relation:
\begin{equation}
\Delta x' = \frac{\rm{1}}{N\Delta f_x} = \Delta x.
\end{equation}
The pixel values in both hologram and image planes are the same via the pixel value $\Delta f_x$ in the intermediate Fourier plane.
Only the zeroth-order region in the intermediate plane, $N\Delta f_x$ is utilized, thereby excluding high-frequency components beyond the diffraction limit. 

Figure 8(a) presents the simulated diffractive waves obtained using the angular spectrum method.
The high-frequency components of diffractions are removed due to an upper bound frequency of $6.25\times10^4 \rm{m}^{-1}$,
effectively functioning as a low-pass filter.
The point image is formed from the primary zone which occupies half the width of the hologram aperture.
The spatial resolution of the point image is 8 $\mu$m, according to the sampling relation.
The spreading angle of the propagating wave is estimated to be 3.9$^\circ$.

The diffraction behavior of a two-fold upsampled hologram was investigated.
The upsampling process increases the pixel resolution from 256$\times$256 pixels at 8 $\mu$m to 512$\times$512 pixels at 4 $\mu$m by duplicating the pixel values. 
The calculation process is performed in the two-fold enhanced Fourier space, based on Eq. (21).
The propagating behavior in Fig. 8(b) demonstrates that the central point image is formed through the entire region of the digital hologram.
The diffracted wave converges at an angle of 7.2$^\circ$ before focusing on the point image, achieving a spatial resolution of 4 $\mu$m.

\begin{figure}[ht!]
\includegraphics[scale=0.75, trim= 0.3cm 11.5cm 0cm 0cm]{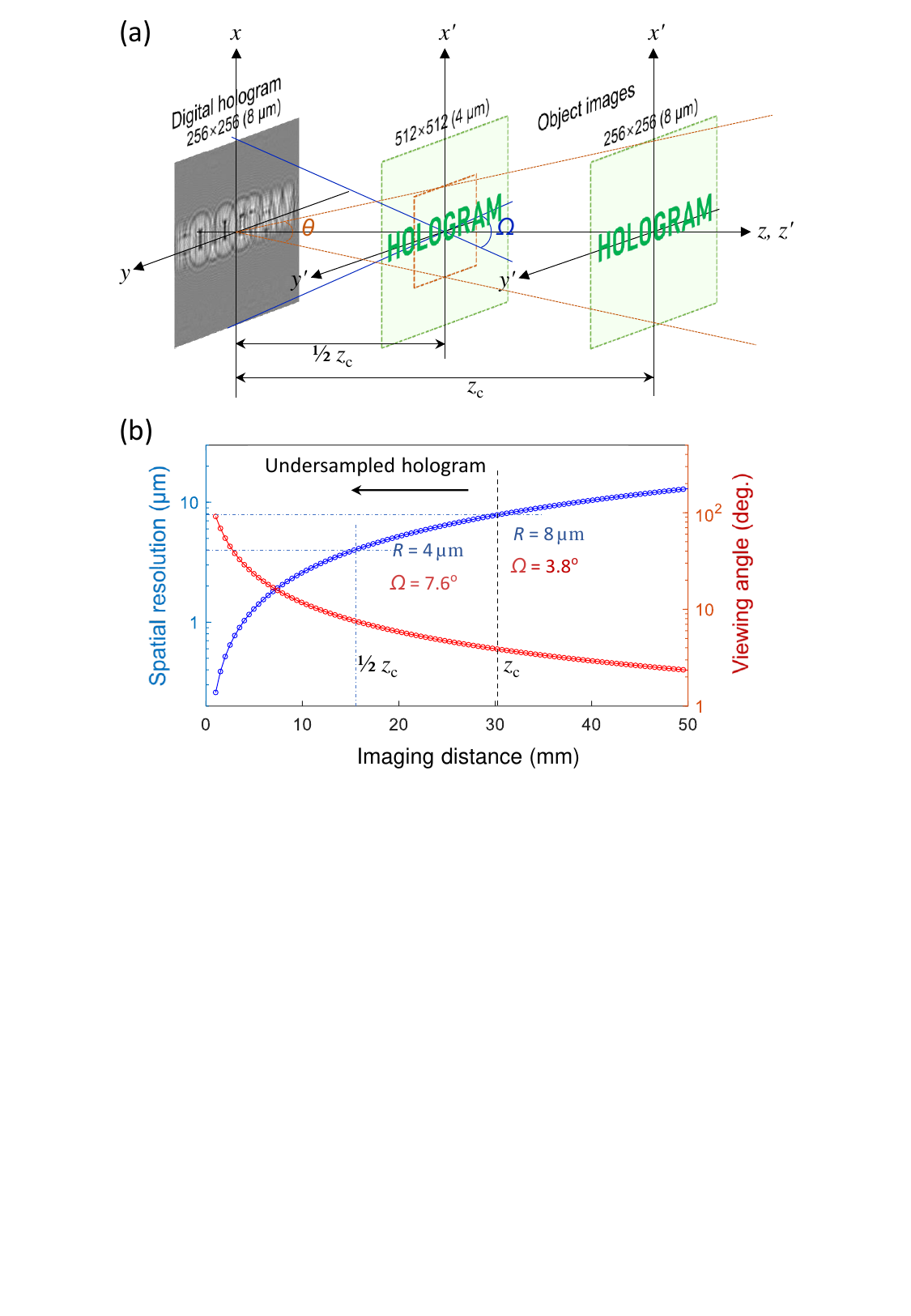}
\caption{Imaging performance of the undersampled hologram.
(a) Geometry of the undersampled hologram and the corresponding restored images.
The restored images from the undersampled hologram exhibit a viewing angle $\it{\Omega}$ greater than the diffraction angle $\theta$ by a hologram pixel.
(b) Variation in spatial resolution and viewing angle of the restored image as a function of imaging distance.}
\end{figure}

This behavior clearly shows that the spatial frequency continuously increases across the shifted replica functions.
Therefore, the maximum spatial frequency in the undersampled hologram is defined by the entire aperture of the hologram in Eq. (1).
Numerical simulation well matches with the theoretical description.
The spatial resolution and viewing angle of the restored image are not fixed from the diffraction angle by a hologram pixel.
It is apparent that these values vary depending on the imaging distance in a digital hologram with finite space-bandwidth, in Fig. 9.
Previously, it was reported that the viewing angle changes according to this relationship, as demonstrated in optical experiments \cite{26}.

\section{Wide-field high-resolution optical imaging}
\subsection{High-resolution image reconstruction from low-resolution hologram}

The image space reconstructed from the undersampled hologram is confined to the diffraction scope by a hologram pixel 
to avoid the interference of high-order images, as displayed in Fig. 9(a).
Meanwhile, the image space satisfying the sampling condition can be expanded to the full extent of the digital hologram.
If the number of pixels increases by a factor of $m$,
the sampling condition in the image plane, $z > \frac{N {\Delta x'}^2}{\lambda}$ is always satisfied because of the $m$-fold reduction in pixel pitch.
For an object space at a half of $z_c$, two-fold expanded space as compared to a proper image space fulfills this sampling condition.  

On the other hand,
the digital hologram suffers from the $m$-fold undersampling, which induces the generation of the high-order images.
As previously described, even under such conditions, the original image can still be restored without losing high-frequency components 
by applying an upsampling process to the undersampled hologram.

Figure 10 shows the image reconsturction behaviour for the digital hologram made from the 2-fold expanded object space at a half of $z_c$.
The digital hologram for the object image consisting of 512$\times$512 pixels with 4-$\mu$m pixel pitch was synthesized 
by using the Rayleigh-Sommerfeld diffraction formula.
The image size of 2,048 $\mu$m equals to the magnitude of digital hologram of 256$\times$256 pixels with 8-$\mu$m pixel pitch.
The reconstructed letter image using the Riemann integral shows overlapping of the $\pm$1$\textsuperscript{st}$-order terms, as seen in Fig. 10(b).
In contrast, the twofold upsampled hologram effectively suppresses these high-order terms,
resulting in a cleaner reconstruction shown in Fig. 10(c).
This reconstruction was performed using the angular spectrum method in Eq. (20), although the Fresnel diffraction method using a single Fourier transform yields similar results.

To evaluate reconstruction fidelity in a realistic scenario, simulations were extended to the fully occupied object composed of both amplitude and phase information.
Standard “pepper” and “boat” images were used to represent the amplitude and phase terms, respectively.
Without upsampling, high-order diffraction images overlap the primary image.
However, the twofold upsampled hologram successfully reconstructs the amplitude and phase images without interference from the $\pm$1$\textsuperscript{st}$-order terms,
as demonstrated in Figs. 10(e) and 10(f).

Figure 11 presents the restoration results for an optically captured hologram.
The complex-amplitude hologram of a USAF resolution target was acquired at a distance of 86.7 mm using off-axis holography.
It was then processed into a final hologram consisting of 540$\times$540 pixels with a 13.8-$\mu$m pixel pitch (see Section 2 of \textcolor{blue}{Supplement 1}).
Given a large critical distance $z_c$=193.3 mm for this configuration, the final hologram is effectively 2.2-fold undersampled.

\begin{figure}[ht!]
\includegraphics[scale=0.8, trim= 1.5cm 19cm 0cm 0cm]{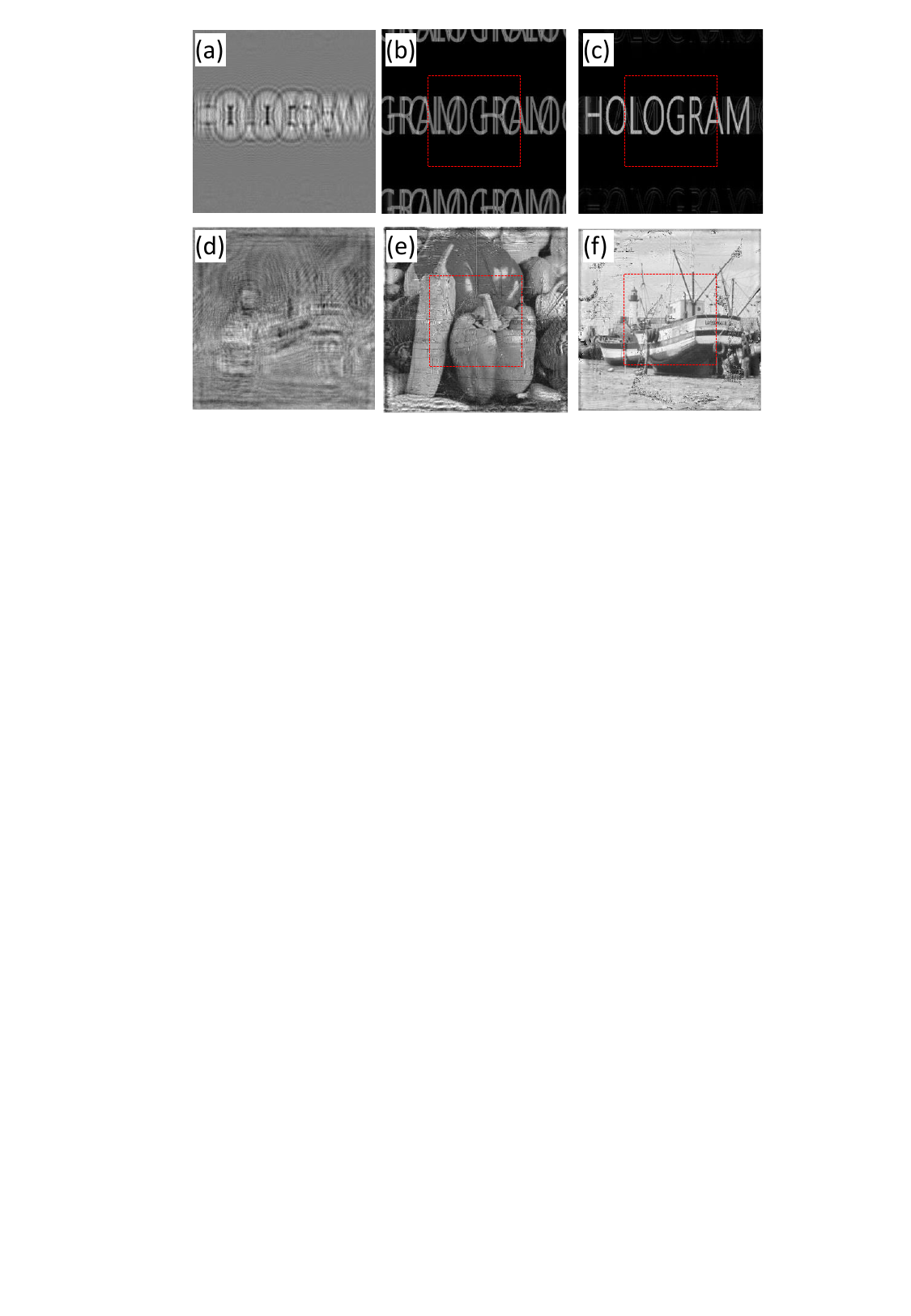}
\caption{Image reconstruction behavior of the undersampled hologram through the removal of high-order noises.
(a) Digital hologram of a letter object, generated from a 2-fold expanded object space at half of $z_c$.
Restored images from the hologram: (b) without and (c) with 2-fold upsampling.
(d) Digital hologram of a fully occupied object.
Restored images of (e) the amplitude and (f) the phase.
Red boxes indicate the diffraction scope defined by a single hologram pixel.}
\end{figure}

\begin{figure}[ht!]
\includegraphics[scale=0.8, trim= 1.5cm 19cm 0cm 0cm]{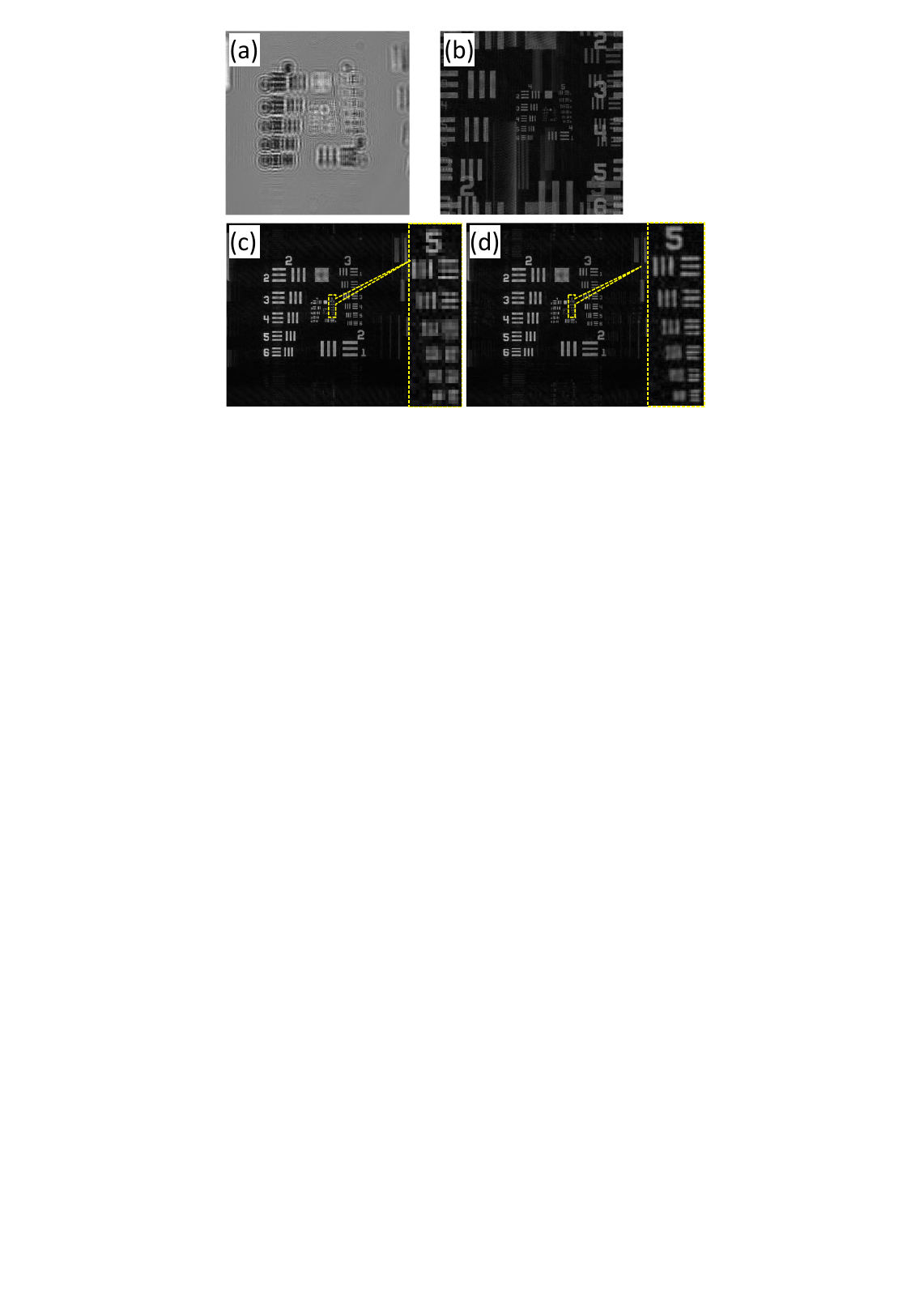}
\caption{Image restoration from an optically captured hologram.
(a) Complex-amplitude hologram of a USAF resolution target.
(b) Reconstructed image using the Fresnel diffraction formula with a single Fourier transform.
Images restored using the angular spectrum method: (c) without and (d) with upsampling.}
\end{figure}

In off-axis holography \cite{2,28,29}, the usable Fourier space of the hologram is limited to half of the full frequency domain.
A plane wave incident on the image sensor at an off-axis angle in the horizontal direction serves as the reference beam.
In this configuration, the Fourier space in the horizontal direction is reduced due to the off-axis geometry,
but the full Fourier space in the vertical direction remains available for reconstruction.

Figure 11(b) shows the reconstructed image obtained using the Fresnel diffraction formula with a single Fourier transform.
The image is contaminated by high-order diffraction terms.
Based on Eq. (1), the spatial resolution determined by the NA of the digital hologram is 6.2 $\mu$m.
Within this resolution range, the line pairs in the vertical direction can be approximately resolved.

Figures 11(c) and 11(d) present the reconstructed images using the angular spectrum method, with and without the upsampling process, respectively.
The magnified view reveals that the 2-fold upsampled hologram successfully recovers the original resolution, particularly in the vertical direction.
However, there is no significant improvement in the horizontal direction due to the loss of Fourier components during the preprocessing of the captured hologram.

\begin{figure}[ht!]
\includegraphics[scale=0.8, trim= 1.2cm 11cm 0cm 0cm]{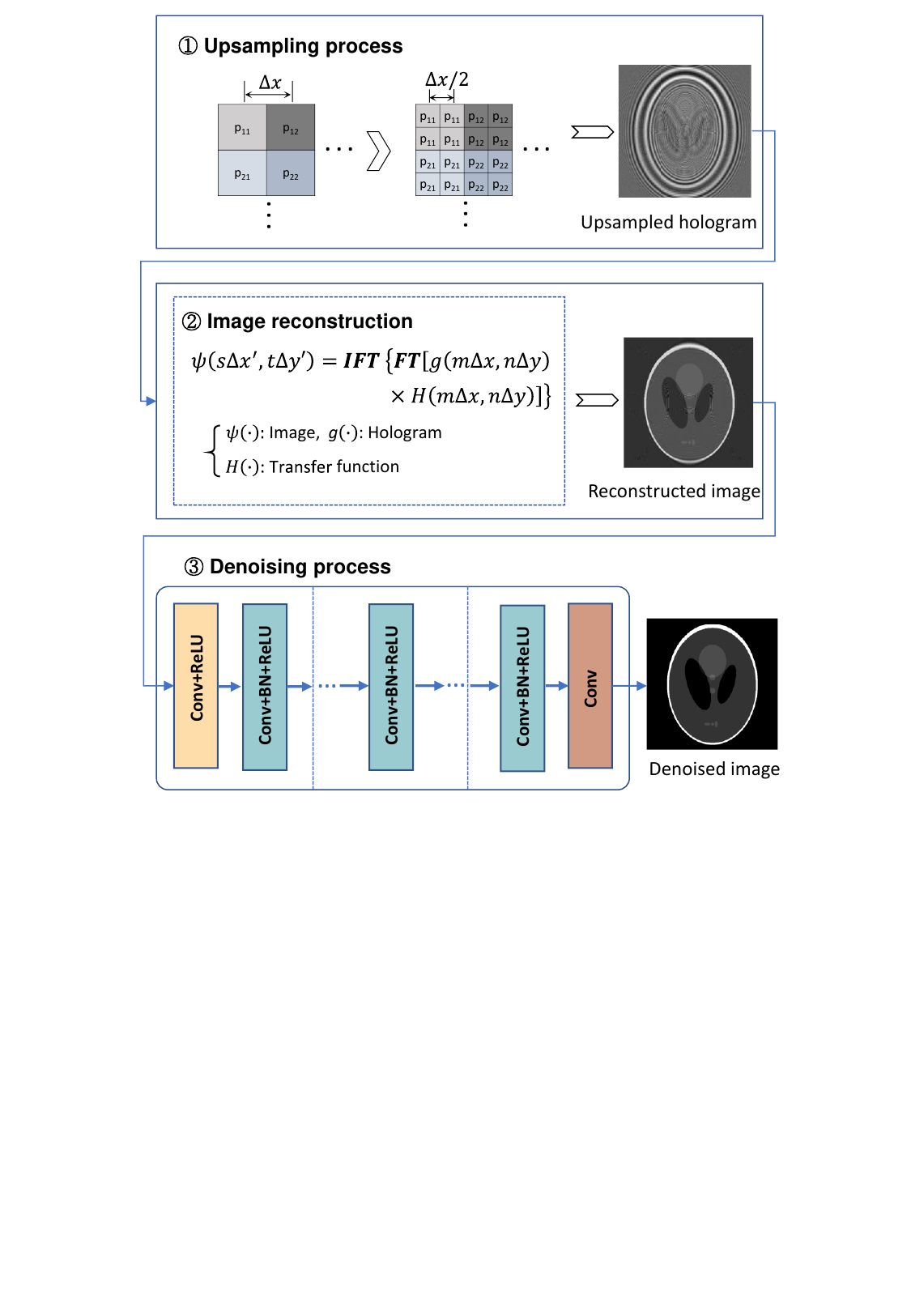}
\caption{Acqusition process of a wide-field high-resolution image from a low-resolution hologram.
A 2-fold upsampling process is shown for clarity.}
\end{figure}

\subsection{Image denoising through learning-based algorithm}

In principle, the upsampled hologram can suppress high-order noise and recover the original pixel information without loss of high-frequency components.
However, as shown in Fig. 6, edge patterns of high-order images may still persist, resulting in residual additive noise in the reconstructed image.
Learning-based algorithms have been shown to effectively remove this specific type of residual noise \cite{30}.

Figure 12 outlines the process of acquiring a wide-field high-resolution image from a low-resolution hologram.
The proposed strategy consists of three main steps:
First, the low-resolution hologram is upsampled by a factor of $m$, according to the desired specification.
Second, the image is reconstructed using the angular spectrum method with an $m$-fold expanded Fourier space.
Third, the resulting noisy image is denoised using a learning-based algorithm.

The denoising convolutional neural network algorithm known as DnCNN was adopted,
which consisting of the feed-forward architecture showing a great performance to remove the additive noises \cite{31}.
The core convolutional layers consist of convolution, batch normlization, and ReLU activation function (see Section 3 of \textcolor{blue}{Supplement 1}).

Figure 13 illustrates the learning-based experiments for denoising the reconstructed noisy images.
The reconstructed noisy images and residual images were used as the input and target data, as displayed in Fig. 13(a).
For calculating the undersampled hologram,
three types of target images as 64$\times$64 pixels with 8-$\mu$m pixel size, 128$\times$128 pixels with 4-$\mu$m pixel size,
and 256$\times$256 pixels with 2-$\mu$m pixel size were prepared by rescaling the handwritten MNIST datasets.
The undersampled holograms for three types of prepared images were calculated at a half of $z_c$, a quarter of $z_c$, and one-eighth of $z_c$, respectively.
All the holograms consist of 32$\times$32 pixels with 8-$\mu$m pixel pitch, and the input noisy images are obtained from the reconstruction of the $m$-fold upsampled holograms.
Each type of dataset contained 3,000 training samples and 1,000 test samples.

The DnCNN network was easily trained and demonstrated strong performance during training (see Section 3 of \textcolor{blue}{Supplement 1}).
After 51 epochs, the trained network accurately restored the original images from the test noisy datasets, as shown in Fig. 13(b).
The peak signal-to-noise ratio (PSNR) values for the restored images from 2-fold, 4-fold, and 8-fold upsampled holograms were approximately 38 dB, 42 dB, and 41 dB, respectively.

\begin{figure}[ht!]
\includegraphics[scale=0.75, trim= 0.7cm 12cm 0cm 0cm]{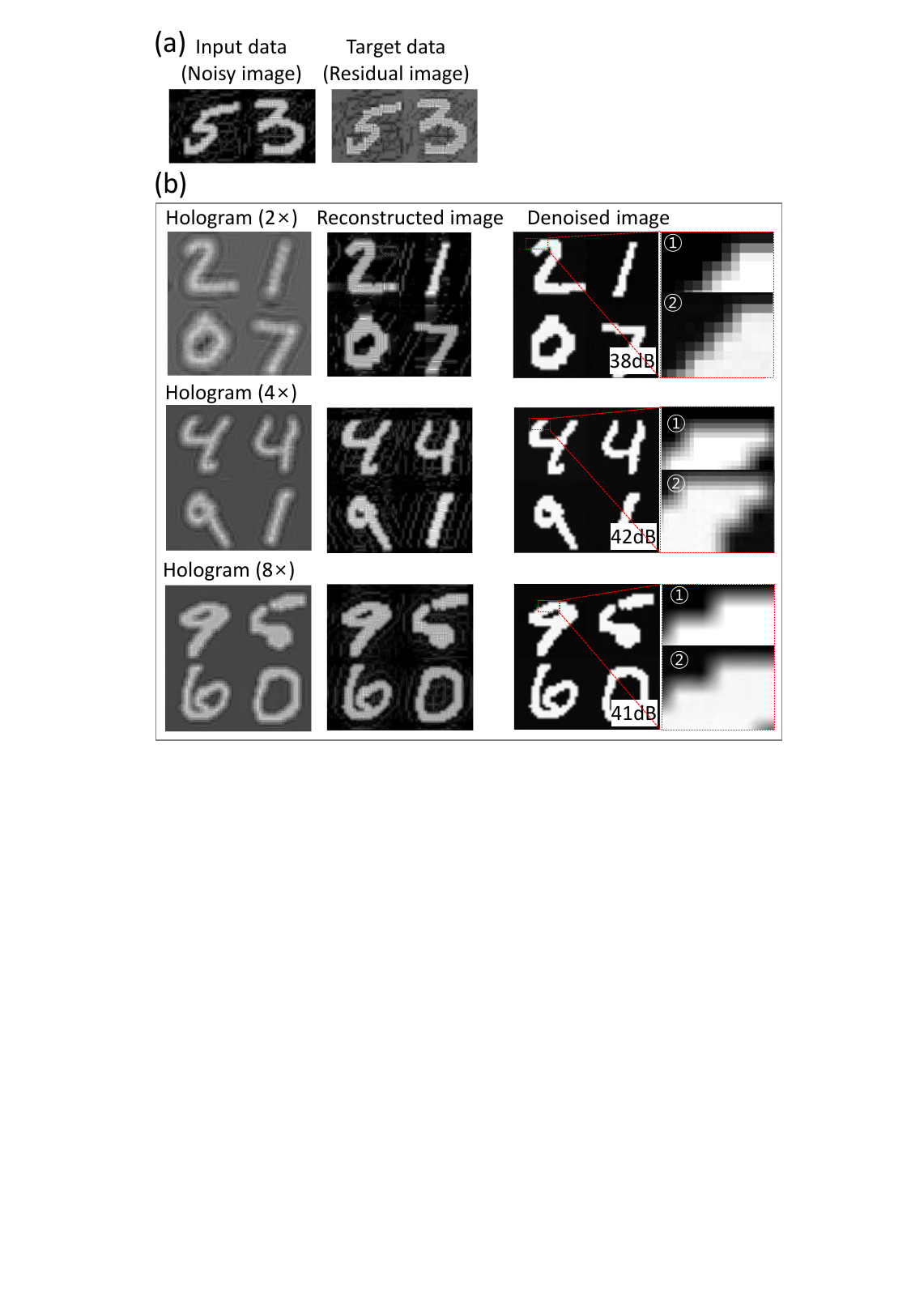}
\caption{Learning-based experiments for denoising reconstructed noisy images.
(a) Input and target MNIST datasets.
(b) Denoising behaviours for restored noisy images from 2-fold, 4-fold, and 8-fold upsampled holograms.
Four sample letter images are shown for convenience.
Magnified denoised images display: $\textcircled{1}$ original image and $\textcircled{2}$ denoised image.}
\end{figure}

Magnified views confirm that the network successfully recovers complete pixel-level details.
For a undersampled hologram generated at one-eighth of $z_c$,
even a low-resolution hologram of 32$\times$32 pixels with 8-$\mu$m pixel pitch could reconstruct a high-resolution image of 256$\times$256 pixels with 1-$\mu$m resolution.
This demonstrates that learning-based algorithms are effective tools for removing the specific noise artifacts that result from residual high-frequency components in high-order diffraction terms.

\section{Discussions}

Wide-field optical microscopy is becoming increasingly important in optical inspection fields such as biomedicine and semiconductor manufacturing \cite{8,11,32,33}.
Conventional optical systems using objective lenses face a fundamental trade-off between spatial resolution and field of view.
Presently, wave-scale resolution is typically achievable only within submillimeter-diameter areas.

Lensless digital holography is free from the limitations imposed by lens performance.
In this system, the space-bandwidth product of the digital hologram dictates both the resolution and the size of the captured image.
In principle, high-resolution imaging over a wide field of view is achievable by capturing a hologram at a short distance using a large image sensor.
However, as discussed earlier, such holograms are severely undersampled due to the large bandwidth required.

The proposed method demonstrates that the captured low-resolution hologram can restore the high-resolution image, despite significant aliasing errors.
This approach is fundamentally different from conventional super-resolution techniques that rely on deep learning or interpolation.
Existing methods typically estimate unknown pixel values through interpolation without access to actually captured high spatial-frequency information.
In contrast, the present method reconstructs a high-resolution image directly from the high-angular-spectrum components encoded in the undersampled hologram,
leading to accurate recovery of image information.
Residual noise resulting from incomplete reconstruction can be effectively addressed using a learning-based denoising algorithm.

A critical requirement of this method is a geometric configuration that allows the complex hologram to be measured at a short distance from the sample.
The beam splitter, used to combine the sample and reference beams, must be as thin as possible to allow acquisition of a high-NA hologram \cite{34}.

Residual noise during the removal of high-order diffraction terms originates from high spatial-frequency components.
Abrupt pixel value changes introduce high-frequency components in the Fourier-transformed data, commonly referred to as ringing artifacts.
Previous studies have shown that these artifacts can be mitigated using window functions or preprocessing techniques in digital holography \cite{35,36,37}.
Further investigation is needed to fully understand the origin of these residual high-frequency components during the suppression of high-order terms.
If appropriate techniques are developed to minimize ringing artifacts, such noise could potentially be eliminated without requiring additional denoising steps.

Furthermore, the concept of spatial-frequency expansion into undersampled regions can be extended to incoherent digital holography.
Incoherent imaging uses low-coherence light, typically a Gaussian wave packet centered around a specific frequency.
Each spectral component of the low-coherence light still obeys the same undersampling principles.
This methodology can also be applied to the Rayleigh–Sommerfeld diffraction regime.
The same replication phenomenon observed in the Fresnel diffraction region can be interpreted on the Ewald sphere (see Section 4 of \textcolor{blue}{Supplement 1}).
Although the replica patterns on the sensor plane appear distorted under these conditions,
they still function as higher spatial-frequency components.

\section{Conclusions}

A mathematical description of the angular spectrum distribution demonstrates that an undersampled hologram at a lower sampling rate can reconstruct an image
with the same spatial resolution and angular field of view as a properly sampled hologram.
The spatial frequencies are found to be continuously distributed across the aliased replica functions.
Moreover, the undersampled replica functions generate high-order diffractive waves that form replica images in the image plane.
When these replica waves are effectively removed through suitable external operations,
high-performance imaging can be achieved, surpassing the inherent space-bandwidth limitations of a digital hologram.
One practical example of this concept addresses the issue of wide-field high-resolution optical imaging.
This technology provides an alternative approach to overcome the finite space-bandwidth constraints of digital holograms, 
paving the way for enhanced imaging performance.

\begin{backmatter}
\bmsection{Funding}
This work was supported by Institute for Information \& Communications Technology Promotion (IITP) grant funded by the Korea government (MSIP) (2021-0-00745)

\bmsection{Disclosures}
The authors declare no conflicts of interest.

\end{backmatter}

\newpage

\section*{\large{Supplementary materials}}

\vspace{3mm}

\setcounter{equation}{0}

\section*{Sampling condition in image reconstruction from the undersampled hologram}

To observe the propagation behavior of the diffracted wave $\psi(x',y')$ from the undersampled hologram $g(x,y)$ to a far distance,
it is essential to first investigate the sampling properties.
The sampling condition in the Fresnel diffraction formula has been well established \cite{1,2,3,4}:
\begin{equation}
\psi(x',y') = \frac{ie^{-i kz_1}}{\lambda z_1} \exp \left[-i\frac{\pi}{\lambda z_1} (x'^2 + y'^2)  \right] \textbf{\textit{FT}} \left[ g(x,y) \exp \left[-i\frac{\pi}{\lambda z_1} (x^2 + y^2)  \right] \right].
\end{equation}

Considering a point-source hologram, the quadratic phase term (QP) within the Fourier transform is expressed as
\begin{equation}
\rm{QP} = \it{} \frac{e^{ikz}}{i \lambda z} \exp \left[ i \frac{\pi}{\lambda} (x^{\rm{2}} + y^{\rm{2}}) \left(\frac{\rm{1}}{z} - \frac{\rm{1}}{z_{\rm{1}}} \right) \right].
\end{equation}
To avoid aliasing, the sampling rate $f_s$ must exceed twice the maximum spatial frequency $f_{x,\rm{max}}$ of the phase term:
\begin{equation}
f_s \geq 2f_{x,\rm{max}} = \frac{2x_{\rm{max}}}{\lambda} \left( \frac{1}{z} - \frac{1}{z_{\rm{1}}} \right).
\end{equation} 
Using a sampling rate of $\frac{1}{\Delta x}$ and the hologram's field size, $N \Delta x = 2|x_{\rm{max}}|$, the well-sampling condition with respect to the reconstruction distance $z_1$ is given by
\begin{equation}
\left \lvert \frac{zz_1}{z_1 - z} \right \rvert \geq z_c =\frac{N {\Delta x}^2} {\lambda}.
\end{equation}

Substituting $z = T z_c$ and $z_1 = S z$, the left-hand tem of Eq. (4) can be rearranged as
\begin{equation}
\left \lvert \frac{z}{1-z/z_1} \right \rvert = \left \lvert \frac{T z_c}{1-1/S} \right \rvert.
\end{equation}
Here, $T$ and $S$ are scaling factors.
Incorporating Eq. (5) into Eq. (4), the sampling condition is represented as follows:
\begin{equation}
\frac{1}{1+T} \leq S \leq \frac{1}{1-T}.
\end{equation}

As illustrated in Fig. 1,
proper sampling is achievable within the intermediate region (I).
High-order diffraction overlapping arises in the region (II), out of the proper sampling range.

Meanwhile, the replica functions preserve the higher spatial frequencies of the original function.
Thus, while aliasing-induced high-order diffractions occur, numerical calculations that retain higher spatial-frequency components are feasible at all propagation distances.

\section*{Optically captured complex hologram using off-axis holography}

An optical hologram was acquired using an off-axis holographic technique, as illustrated in Fig. 2(a).
A collimated laser beam was split into an object wave and a reference wave using a beam splitter.
A 532-nm green laser (LCX-532, Oxxius) served as the light source.
The object beam illuminated a negative USAF resolution target, and the diffracted waves from the target were recorded on an image sensor.
The reference wave, incident on the sensor at an off-axis angle, formed interference fringes.
A digital hologram with 2160$\times$2160 pixels of a 3.45-$\mu$m pixel pitch was captured using an image sensor (GS3-U3-89S6M-C, FLIR).

The complex-amplitude hologram was obtained by isolating the first-order region of the Fourier-transformed hologram data.
As shown in Fig. 2(c), the reference plane wave was directed at the sensor at an off-axis angle along the horizontal axis.
To eliminate the non-diffracted (DC) terms and twin image artifacts, a significant portion of the Fourier space in the horizontal direction was discarded, while the vertical direction remained mostly unaffected.
The cropped region was then shifted to the center, and an inverse Fourier transform was applied to generate the final complex-amplitude hologram.

\section*{Architecture of DnCNN networks and its training performance}

Figure 3(a) illustrates the feed-forward architecture of the DnCNN network \cite{5}.
The network consists of 17 convolutional layers, where the main convolutional block comprises three operations: convolution, batch normalization, and ReLU activation.
Each convolutional layer contains 64 filters with a kernel size of 3$\times$3.
A stride and padding of 1 are applied to maintain the same input and output dimensions across all layers.
All convolutional layers output 64 channels, representing different feature maps learned at each stage.

Figure 3(b) shows the setup for generating undersampled holograms and the corresponding images used in training dataset preparation.
First, the handwritten MNIST dataset was resized into three different image types--64$\times$64, 128$\times$128, and 256$\times$256 pixels--using interpolation.
Each dataset contained 3,000 training samples and 1,000 test samples.
Undersampled holograms with a resolution of 32$\times$32 pixels and an 8-$\mu$m pixel pitch were computed using the Riemann integral of the Rayleigh-Sommerfeld diffraction formula at distances of half, one-quarter, and one-eighth of the critical distance ($z_c$), which is calculated to be 3.85 mm.

Next, three types of input noisy images were reconstructed from the upsampled holograms.
For example, at one-quarter of $z_c$, images with a resolution of 128$\times$128 pixels and a 4-$\mu$m pixel pitch were reconstructed from 4-fold upsampled holograms.
The angular spectrum method was used for image reconstruction.

The image denoising is modeled as follows:
\begin{equation}
y = x +v,
\end{equation}
where $y$ is the input noisy image, $x$ is the clean rescaled MNIST image, and $v$ is the residual image.
The residual images, obtained by subtracting the clean images from the noisy inputs, are used as target data during training.

Figure 4 shows the evolution of the loss value as a function of training epochs.
The mean squared error (MSE) was used as the loss function for regression.
The loss function $L(w)$, which measures the difference between the predicted residual $R(y)$ and the actual residual $v$, is defined as:
\begin{equation}
L(w) = \frac{1}{N} \sum_{i=1}^{N} \parallel R(y_i ;w) - v_i \parallel^2,
\end{equation}
where $w$ represents the weights of the convolutional filters.
These weights are updated at each iteration using the Adam optimization algorithm, with a learning rate set to $10^{-4}$.
The network demonstrates strong convergence performance across all three types of training datasets.
The loss rapidly decreases and stabilizes after only a few training epochs.

\section*{Undersampled hologram in the Rayleigh-Sommerfeld diffraction region}

Figure 5(a) illustrates the geometry for capturing a digital hologram in the Rayleigh–Sommerfeld diffraction region.
The wide-angle diffraction behavior is well represented on the Ewald sphere \cite{6,7}.
The Rayleigh-Sommerfeld diffraction uses a kernel of $\frac{1}{r} e^{ikr}$,
where $r=\sqrt{x^2+y^2+z^2}$ defines the radius of the Ewald sphere.

The lateral spatial frequencies along the $x$ and $y$ axes are given by:
\begin{equation}
f_x = \frac{x}{\lambda r} ,\;  f_y = \frac{y}{\lambda r}.
\end{equation}
In an undersampled hologram at a lower sampling rate, replica patterns appear clearly along the Ewald sphere,
which can be interpreted as phase modulation caused by integer multiples of the sampling frequency.
However, the diffraction patterns on the sensor plane appear distorted, as shown in Fig. 5(a).

Figure 5(b) shows an example of an undersampled hologram generated in the Rayleigh–Sommerfeld regime.
The hologram, consisting of 256$\times$256 pixels with a 2-$\mu$m pixel pitch, was recorded at one-eighth of the critical distance $z_c$,
yielding a spatial resolution of 0.25 $\mu$m for a point object.
When the spatial resolution exceeds the wavelength scale, noticeable distortion arises, exhibiting a pincushion-like shape.
Despite the severe distortion, it is evident that a high-resolution point image can still be reconstructed through the inverse process.

\renewcommand\thefigure{\arabic{figure}}
\setcounter{figure}{0}

\begin{figure}[ht!]
\includegraphics[scale=0.85, trim=1.8cm 19.8cm 0cm -1cm]{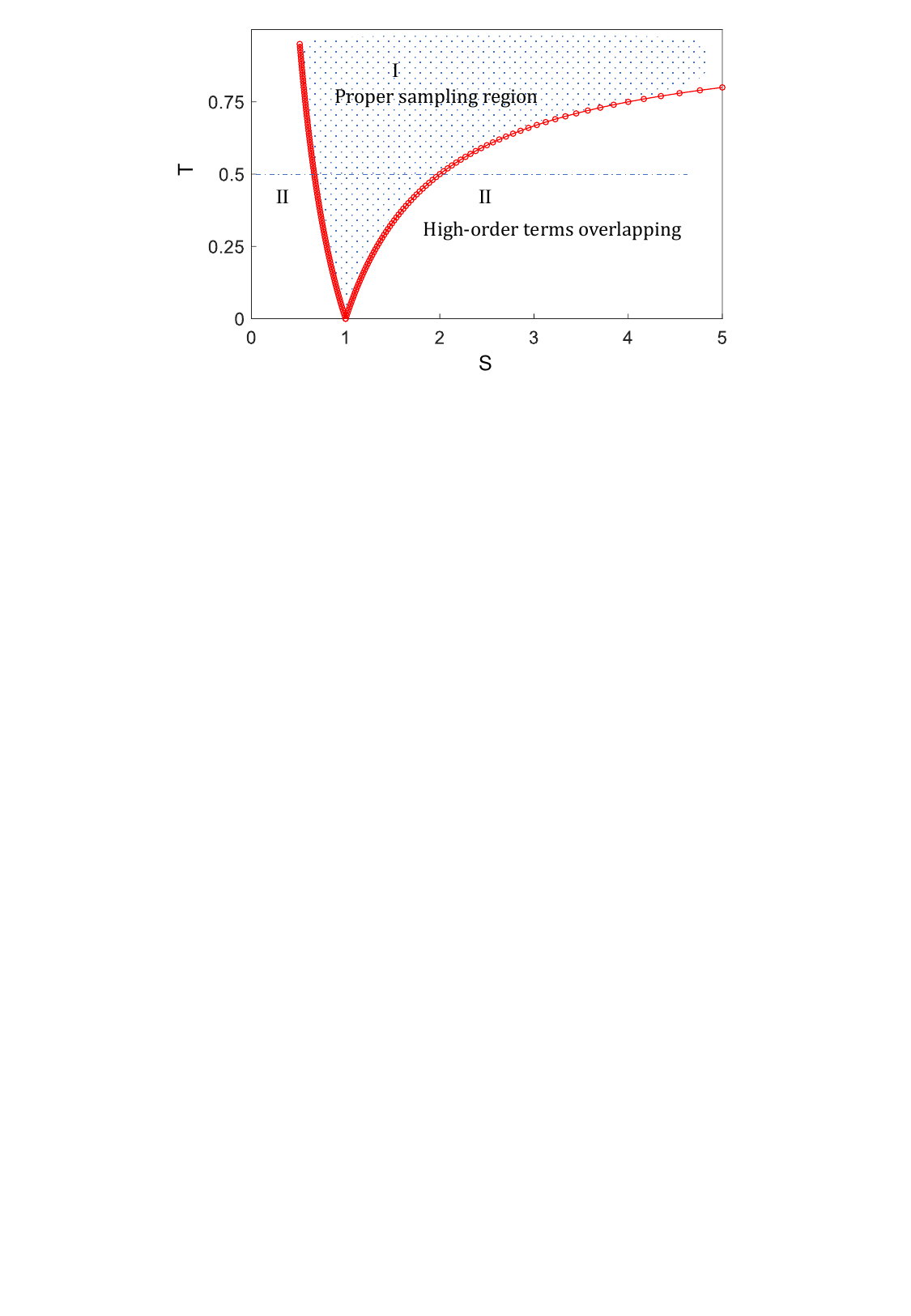}
\caption{Sampling conditions in numerical simulation of Fresnel diffraction for the undersampled hologram.
Numerical calculations conserving higher spatial-frequency components can be performed at all propagation distances.}
\end{figure}

\begin{figure}[ht!]
\centering\includegraphics[scale=0.75, trim= 1.5cm 8cm 1cm 0cm]{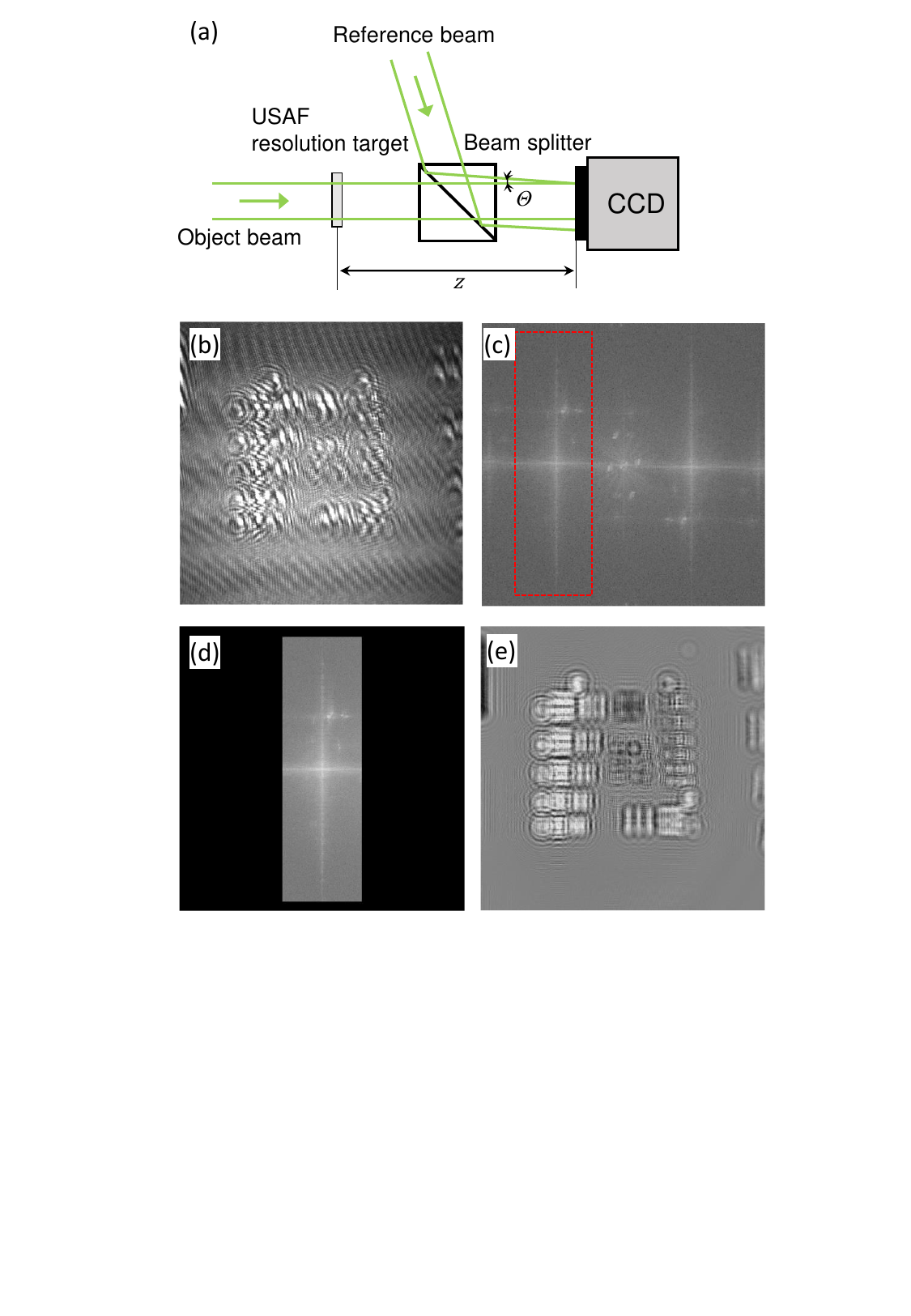}
\caption{(a) Schematics of off-axis holographic technique. (b) Optical hologram for the USAF resolution target
and (c) Fourier-transformed data of the hologram. Red box indicates the cropped first-order region. 
(d) Cropped region shifted to the center. (e) Final complex-amplitude hologram. $\it{\Theta}$ is the off-axis angle.}
\end{figure}

\begin{figure}[ht!]
\includegraphics[scale=0.8, trim= 1.5cm 10cm 0cm 0cm]{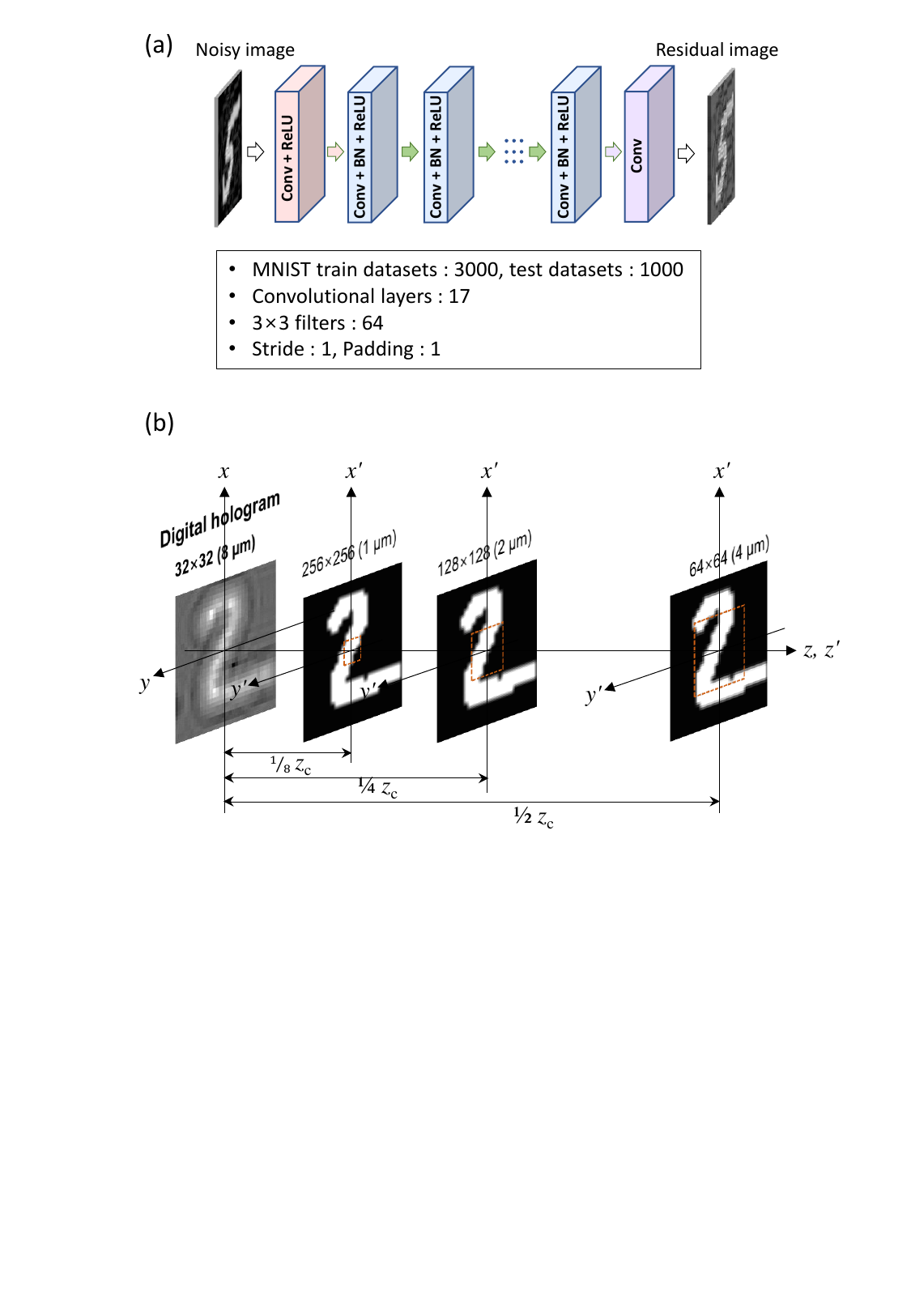}
\caption{ (a) Feed-forward architecture of the DnCNN network.
(b) Geometry of the undersampled hologram and corresponding images used to prepare the training datasets.
For illustration, the digit "2" is shown.}
\end{figure}

\begin{figure}[ht!]
\includegraphics[scale=0.85, trim= 1.7cm 19.5cm 0cm 0cm]{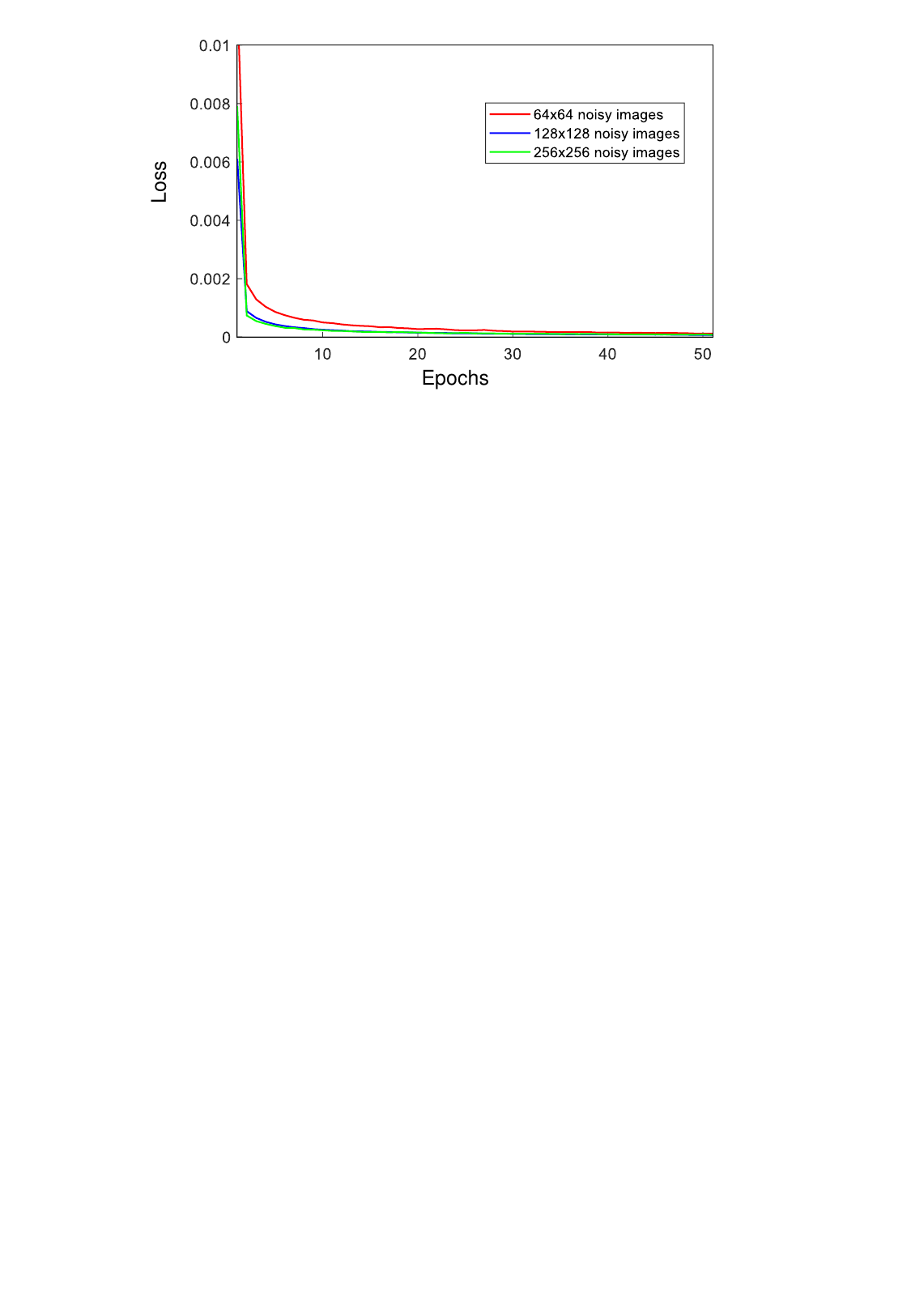}
\caption{Change in the loss value as a function of training epochs for the DnCNN network.}
\end{figure}

\begin{figure}[ht!]
\includegraphics[scale=0.8, trim= 1.5cm 10cm 0cm 0cm]{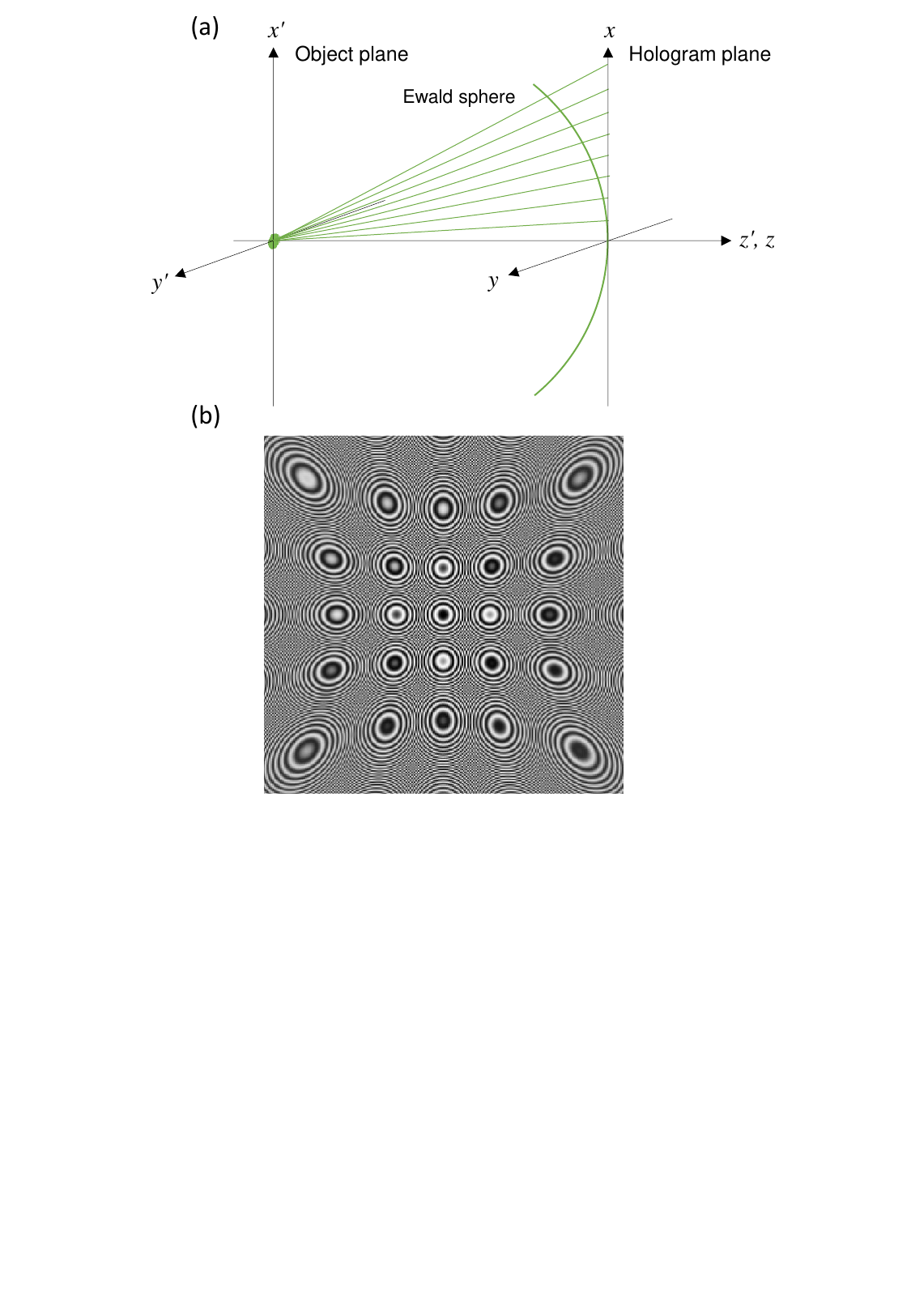}
\caption{ (a) Geometry for capturing a digital hologram in the Rayleigh-Sommerfeld diffraction region.
(b) The undersampled hologram for point object of 0.25-$\mu$m resolution located at one-eighth of $z_c$.}
\end{figure}

\end{document}